\documentclass[12pt]{article}
\usepackage[english]{babel}
\usepackage{graphicx}
\usepackage{epsfig}
\usepackage{color}

\newcommand{\beq}      {\begin{eqnarray}}
\newcommand{\eeq}      {\end{eqnarray}}

\newcommand{\bq}       {\begin{eqnarray}}
\newcommand{\eq}       {\end{eqnarray}}

\title
{An exploratory analysis of the transient and long term behavior  of  small 3D perturbations
in the circular cylinder wake\\
}

\author{S. Scarsoglio$^\sharp$, D. Tordella$^{\sharp, *}$ and W. O. Criminale$^\flat$}
\date{}

\begin{document}
\maketitle

{\it $^\sharp$  Dipartimento di Ingegneria Aeronautica e Spaziale,
Politecnico di Torino,  \textit{10129 Torino, Italy},

\it $^\flat$ Department of Applied Mathematics, University of
Washington, Seattle,  \textit{WA 98195-2420, USA}}

\begin{abstract}
An initial-value problem (IVP) for arbitrary small
three-dimensional vorticity perturbations imposed on a free shear
flow is
considered. 
The viscous perturbation equations are then combined in terms of
the vorticity and velocity, and are solved by means of a combined
Laplace-Fourier transform in the plane normal to the basic flow.
The perturbations can be uniform or damped along the mean flow
direction. This treatment allows for a simplification of the
governing equations such that it is possible to observe long
transients, that can last hundreds time scales. This result would
not be possible  over an acceptable lapse of time
 by carrying out a direct numerical integration of
the linearized Navier-Stokes equations. The exploration is done
with respect to physical inputs as the angle of obliquity, the
symmetry of the perturbation and the streamwise damping rate. The
base flow is an intermediate section of the growing
two-dimensional circular cylinder wake where the entrainment
process is still active. Two Reynolds numbers of the order of the
critical value for the onset of the first instability are
considered. The early transient evolution offers very different
scenarios for which we present a summary for particular cases. For
example, for amplified perturbations, we have observed two kinds
of transients, namely (1) a monotone amplification and (2) a
sequence of growth - decrease - final growth. In the latter case,
if the initial condition is an asymmetric oblique or longitudinal
perturbation, the transient clearly shows an initial oscillatory
time scale. That increases moving downstream, and is different
from the asymptotic value. Two periodic temporal patterns are thus
present in the system. Furthermore, the more a perturbation is
longitudinally confined the more it is amplified in time.  The
long-term behavior of two-dimensional disturbances shows
excellent agreement with a recent two-dimensional spatio-temporal
multiscale modal analysis and with laboratory data concerning the
frequency and wave length of the parallel vortex shedding in the
cylinder wake.

{\bf Keywords:} {Initial-value problem (IVP) - perturbation -
early transient - spatially developing flows -instability -
absolute - periodic temporal patterns - growth -
three-dimensional}.

{\bf * corresponding author, e-mail: daniela.tordella@polito.it}
\end{abstract}

\section{Introduction}

 Recent shear flows studies (\cite{BF92}, \cite{CD90},
\cite{CLZ91}) have shown the importance of the early time
dynamics, that in principle can lead to non-linear growth long
before an exponential mode is dominant. The recognition of the
existence of an algebraic growth, due - among other reasons - to
the non-orthogonality of the eigenfunctions (\cite{S49}) and a
possible resonance between Orr-Sommerfeld and Squire solutions
(\cite{BG81}), recently promoted many contributions directed to
study the early-period dynamics. For fully bounded flows works by
\cite{CLZ91}, \cite{CJLJ97}, \cite{G91}, \cite{B93}, \cite{SH94},
\cite{S2007}, and for partially bounded flows works by
\cite{LJJC99}, \cite{HG81}, \cite{CB00}, \cite{CD00} can be cited. As for free
shear flows, the attention was first aimed in order to obtain
closed-form solutions to the initial-value inviscid problem
(\cite{BC94}, \cite{CJL95}) and was successful by considering
piecewise linear parallel basic flow profiles.

An interesting aspect observed in the intermediate periods
is that the maximal amplification is generally associated to oblique disturbances, that, as a consequence, potentially can promote early transition, see e.g.
\cite{CB00}. In fact, these perturbations, which are asymptotically stable at all Reynolds numbers, are the perturbations  best exploiting the energy transient amplification.

In this work we consider as a prototype for free shear flow the
two-dimensional wake past a bluff body. The wake stability has
been widely studied by means of modal analyses (e.g. \cite{MC72},
\cite{TTC86}, \cite{HA87}, \cite{HM90}). However, in this way only
the asymptotic fate can be determined, regardless of the transient
behavior and the underlying physical cause of any instability.

In this work we adopt the velocity-vorticity formulation  to
evaluate the general initial-value perturbative problem. This
method was proposed by Criminale,  Drazin and co-authors in the
years 1990-2000 (\cite{CD90}, \cite{CD00}, \cite{CJL95},
\cite{CJLJ97}, \cite{LJJC99}). In synthesis, the variables are
Laplace-Fourier transformed in the plane normal to the basic flow.
Afterward, the resulting partial differential equations in time
are integrated numerically. This procedure allows for completely
arbitrary initial expansions by using a known set of functions
(Schauder basis in the $L^2$ space) and yields the complete
dynamics
--- the early time transients and the asymptotic behavior (up to many hundred time scales) --- for any disturbance.
The long term dynamics would not, in fact, have been easily
recovered by using the alternative method of the direct numerical
integration of the linear equations since the integration over a
range of time larger than a few dozen basic time scales is not
feasible.

The base flow model that we employ includes the wake transversal
velocity and thus the nonlinear and diffusive dynamics that are
responsible for the growth of the flow and the associated mass
entrainment. We consider the first two order terms of the
analytical Navier-Stokes expansion obtained by Tordella and Belan
(2002, 2003) (\cite{BT02}, \cite{TB03}), see \S \ref{Base flow}.
In particular, we consider the longitudinal component of such an
expansion solution, the problem is parameterized by $x_0$, the
longitudinal coordinate, and the Reynolds number $Re$.

We use a complex wavenumber for the disturbance component aligned
with the flow so that longitudinal spatially damped waves are
represented. It should be observed that a longitudinal spatial
growth could not be considered physically admissible as an initial
condition since the energy density of the initial perturbation
would be infinite. In the context of the initial-value problem,
this is an innovative feature adopted to introduce a possible
spatial evolution (damping) of the perturbative wave in the
longitudinal direction. The perturbative equations are numerically
integrated by the method of lines. The equations formulation and
initial and boundary conditions are presented in \S
\ref{Formulation}.

The perturbation evolution is examined for the base flow
configurations corresponding to the Reynolds numbers of $50, 100$,
and for a typical section, $x_0 = 10$, of the intermediate region
of the flow where the entrainment process is active. A comparison
with a base flow far field configuration, $x_0 = 50$, is also
proposed. The normalized perturbation kinetic energy density is
the physical quantity on which the transient growth is observed
(see \S \ref{Transient}). To determine the temporal asymptotics of
the disturbance an equivalent of the modal temporal growth rate is
introduced (see \S \ref{Asymptotic}).

\noindent In the case of longitudinal disturbances,  comparison
with recent spatio-temporal multiscale Orr-Sommerfeld analysis
(\cite{BT06}, \cite{TSB06}) and with laboratory experimental data
(\cite{W89}) is carried out, see \S \ref{Asymptotic}. As noted,
the initial conditions posed are arbitrary. As far as the modal
theory is concerned, the agreement is excellent both for the
frequency, defined as the temporal derivative of the perturbation
phase, and the temporal growth rate. It is also in quantitative
agreement with respect to the laboratory data. The experiment
shows couples of pulsation and wavelength of the cylinder vortex
shedding that are close to those yielded by the IVP analysis when
the wavenumber is that where the growth rate is maximum.
Conclusions are given in \S \ref{Conclusions}.

\section{Initial-value problem}

\subsection{Base flow}\label{Base flow}

The base flow is considered viscous and incompressible. To
describe the two-dimensional growing wake flow, an expansion
solution for the Navier-Stokes two-dimensional steady bluff body
wake (\cite{TB03}, \cite{BT02}) has been used. The $x$ coordinate
is parallel to the free stream velocity, the $y$ coordinate is
normal. This approximated analytical Navier-Stokes solution
incorporates the effects due to the full non linear convection as
well as the streamwise and transverse diffusion. The solution was
obtained by matching an inner Navier-Stokes expansion in terms of
the inverse of the longitudinal coordinate $x$ ($x^{-n/2}$, $n =
0, 1, 2, \ldots$) with an outer Navier-Stokes expansion in terms
of the inverse of the distance from the body.

\begin{figure}
  \centerline{\includegraphics[height=8cm,width=9cm]{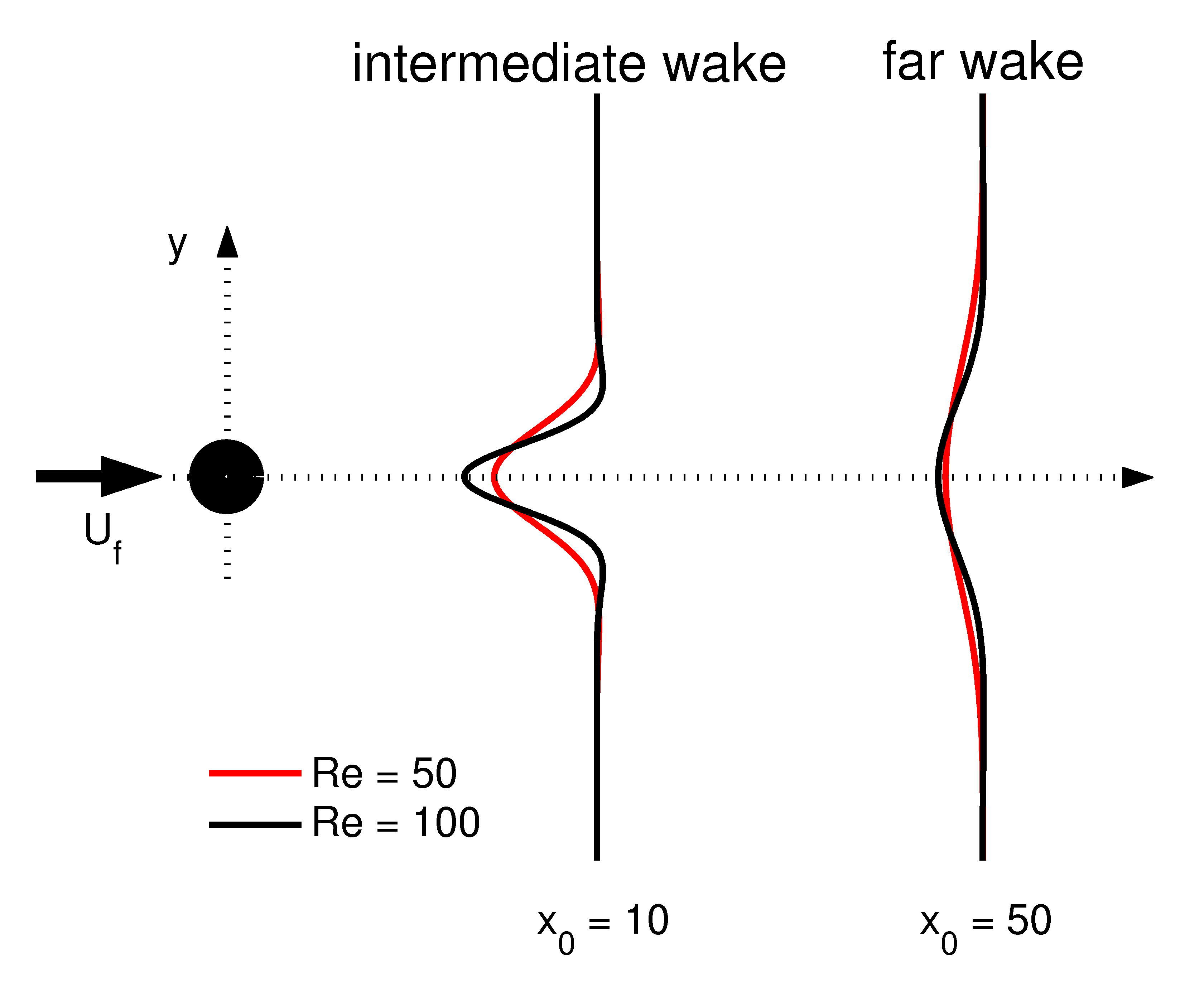}}
  \caption{Wake schematic. Profile $U_f(y; x_0, Re)$ in the intermediate ($x_0 = 10$) and far ($x_0 = 50$)
  wake for different Reynolds numbers, $U_f$ is the free stream velocity. The diameter of the cylinder is out of scale (three times)
  with respect to the wake profiles.}\label{fig:wake_profile}
\end{figure}

\noindent Here we take the first two orders ($n = 0, 1$) of the
inner longitudinal component of the velocity field as a first
approximation of the primary flow. In the present formulation the
near-parallel hypothesis for the base flow, at a longitudinal
position $x = x_0$, is made. The coordinate $x_0$ plays the role
of parameter of the steady system together with the Reynolds
number. The analytical expression for the profile of the
longitudinal component is
\begin{eqnarray}
\noindent  U(y; x_0, Re)=1 - a C_1 x_0^{-1/2} \rm
e^{\displaystyle{- \frac {Re}{ 4} \frac {y^2}{ x_0}}}
\label{U_wake_profile}
\end{eqnarray}
\noindent where $a$ is related to the drag coefficient ($a =
\displaystyle{\frac 1 4} (Re / \pi)^{1/2} c_D(Re)$) and $C_1$ is an integration constant depending on the Reynolds number.
As said in the introduction, this two terms representation is extracted from an analytical asymptotic expansion
where the velocity vector and the pressure are determined to the forth order. It should be observed that the transversal velocity  component
$V$ first appears at the third order $(n=2)$, while the pressure only at the forth order $(n=3)$.
Up to the second order, the field is thus parallel. Beyond the second order the analytical expression becomes much more complex, special functions as the confluent hypergeometric functions plays a role
(\cite{BT02}) associated to the deviation from parallelism. By changing the $x_0$ values, the base flow profile
(\ref{U_wake_profile}) will locally approximate the behavior of
the actual wake generated by the body. Here, the region
considered, if not otherwise specified, is fixed to a typical
section, $x_0 = 10 D$ (where $D$ is the spatial scale of the wake)
of the intermediate wake. The term intermediate is used in the
general sense as used by \cite{B66}: 'intermediate asymptotics are
self-similar or near-similar solutions of general problems, valid
for times, and distances from boundaries, large enough for the
influence of the fine details of the initial/or boundary
conditions to disappear, but small enough that the system is far
from the ultimate equilibrium state...'. The distance beyond which
the intermediate region is assumed to begin varies from eight to
four diameters $D$ for $Re \in [20, 40]$ (see \cite{BT02},
\cite{TB03}). Base flow configurations corresponding to a $Re$ of
$50, 100$ are considered. In Figure \ref{fig:wake_profile} a
representation of the wake profile at differing longitudinal
stations is shown.

\subsection{Formulation}\label{Formulation}

By exciting the base flow with small arbitrary three-dimensional
perturbations, the continuity and Navier-Stokes equations that
describe the perturbed system are
\begin{equation}
\frac{\partial \widetilde{u}}{\partial x} + \frac{\partial
\widetilde{v}}{\partial y} + \frac{\partial
\widetilde{w}}{\partial z} = 0 \label{continuity}
\end{equation}
\begin{equation}
\frac{\partial \widetilde{u}}{\partial t} + U \frac{\partial
\widetilde{u}}{\partial x} + \widetilde{v} \frac{\partial
U}{\partial y} +  \frac{\partial \widetilde{p}}{\partial x} =
\frac{1}{Re} \nabla^2\widetilde{u} \label{NS1}
\end{equation}
\begin{equation}
\frac{\partial \widetilde{v}}{\partial t} + U \frac{\partial
\widetilde{v}}{\partial x} + \frac{\partial
\widetilde{p}}{\partial y} = \frac{1}{Re} \nabla^2\widetilde{v}
\label{NS2}
\end{equation}
\begin{equation}
\frac{\partial \widetilde{w}}{\partial t} + U \frac{\partial
\widetilde{w}}{\partial x} + \frac{\partial
\widetilde{p}}{\partial z} = \frac{1}{Re} \nabla^2\widetilde{w}
\label{NS3}
\end{equation}
\noindent where ($\widetilde{u}(x, y, z, t)$, $\widetilde{v}(x, y,
z, t)$, $\widetilde{w}(x, y, z, t)$) and $\widetilde{p}(x, y, z,
t)$ are the perturbation velocity components and pressure
respectively. The independent spatial variables $z$ and $y$ are
defined from $-\infty$ to $+\infty$, while  $x$ is defined in the semispace occupied by the wake, from $0$ to $+\infty$.
All physical quantities are normalized with respect to the free
stream velocity $U_f$, the body scale $D$ and the
density. \noindent By combining equations (\ref{NS1}) to
(\ref{NS3}) to eliminate the pressure, the linearized equations
describing the perturbation dynamics become
\begin{equation}
(\frac{\partial }{\partial t} + U \frac{\partial }{\partial x})
\nabla^2 \widetilde{v} - \frac{\partial \widetilde{v}}{\partial
x}\frac{d^2U}{dy^2} = \frac{1}{Re} \nabla^4 \widetilde{v}
\label{OS}
\end{equation}
\begin{equation}
(\frac{\partial }{\partial t} + U \frac{\partial }{\partial
x})\widetilde{\omega}_y + \frac{\partial \widetilde{v}}{\partial
z}\frac{dU}{dy} = \frac{1}{Re} \nabla^2 \widetilde{\omega}_y
\label{SQUIRE}
\end{equation}
\noindent where $\widetilde{\omega}_y$ is the transversal
component of the perturbation vorticity field. By introducing the
quantity $\widetilde{\Gamma}$, that is defined by
\begin{equation}
\nabla^2\widetilde{v} = \widetilde{\Gamma} \label{Vel_Vor}
\end{equation}
\noindent we obtain three coupled equations (\ref{OS}),
(\ref{SQUIRE}) and (\ref{Vel_Vor}). Equations (\ref{OS}) and
(\ref{SQUIRE}) are the Orr-Sommerfeld and Squire equations
respectively, from the classical linear stability analysis for
three-dimensional disturbances. From kinematics, the relation
\begin{equation}
\widetilde{\Gamma} = \frac{\partial \widetilde{\omega}_z}{\partial
x} - \frac{\partial \widetilde{\omega}_x}{\partial z}
\label{Vel_Vor_Kin}
\end{equation}
\noindent physically links together the perturbation vorticity
components in the $x$ and $z$ directions ($\widetilde{\omega}_x$
and $\widetilde{\omega}_z$ respectively) and the perturbed
velocity field. By combining equations (\ref{OS}) and
(\ref{Vel_Vor}) then
\begin{equation}
\frac{\partial \widetilde{\Gamma}}{\partial t} + U \frac{\partial
\widetilde{\Gamma}}{\partial x} - \frac{\partial
\widetilde{v}}{\partial x}\frac{d^2U}{dy^2} = \frac{1}{Re}
\nabla^2 \widetilde{\Gamma} \label{OS_Gamma}
\end{equation}
which, together with (\ref{SQUIRE}) and (\ref{Vel_Vor}), fully
describes the perturbed system in terms of vorticity.  This
formulation is a classical one. Alternative classical formulations,
as the velocity-pressure one, are in common use. We chose this formulation because
the vorticity transport and diffusion is the principal phenomenology for
the dynamics of a wake system. For piecewise linear profiles for $U$
analytical solutions can be found. For continuous profiles, the
governing perturbative equations cannot be analytically solved in
general, but may assume a reduced form in the free shear case
(\cite{CJJ03}).

\noindent Moreover, from the equations (\ref{SQUIRE}),
(\ref{Vel_Vor}) and (\ref{OS_Gamma}), it is clear that the
interaction of the mean vorticity in $z$-direction ($\Omega_z = -
dU/dy$) with the perturbation strain rates in $x$ and $z$
directions ($\frac{\partial \widetilde{v}}{\partial x}$ and
$\frac{\partial \widetilde{v}}{\partial z}$ respectively) proves
to be a major source of any perturbation vorticity production.

\begin{figure}[t]
  \center
  \includegraphics{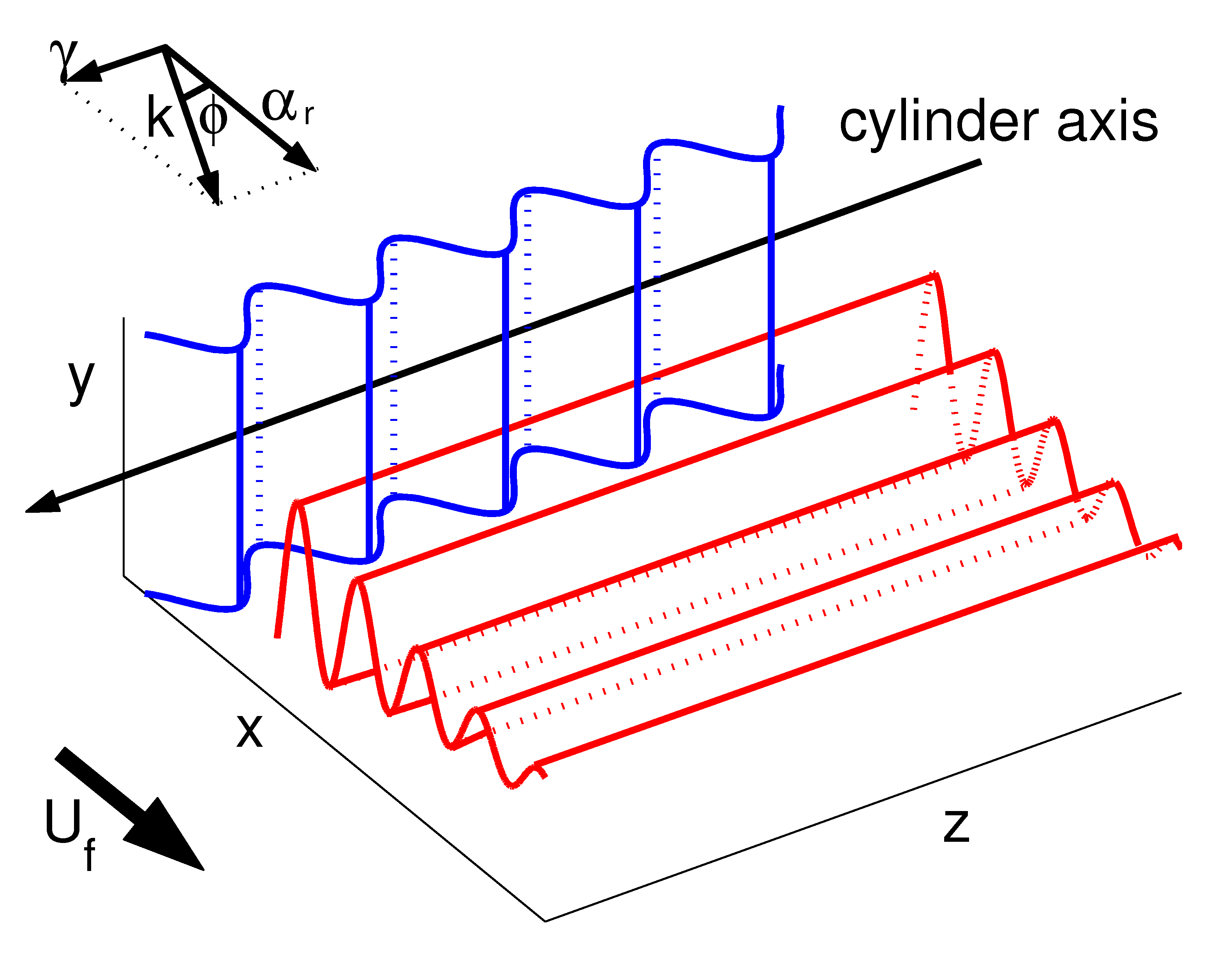}\\
  \caption{Perturbation geometry scheme.}\label{DT_perturbation_scheme}
\end{figure}

The perturbation quantities are Laplace and Fourier decomposed in
the $x$ and $z$ directions, respectively. A complex wavenumber
$\alpha = \alpha_r + i \alpha_i$ along the $x$ coordinate as well
as a real wavenumber $\gamma$ along the $z$ coordinate are
introduced. In order to have a finite perturbation kinetic energy,
the imaginary part $\alpha_i$ of the complex longitudinal
wavenumber can only assume non-negative values. In so doing, we
allow for perturbative waves that can spatially decay ($\alpha_i >
0$) or remain constant in amplitude ($\alpha_i = 0$). The
perturbation quantities $(\widetilde{v}, \widetilde{\Gamma},
\widetilde{\omega}_y)$ involved in the system dynamics are now
indicated as $(\hat{v}, \hat{\Gamma}, \hat{\omega}_y)$, where

\begin{equation}\label{IVP2_Four_trans}
\hat{g}(y, t; \alpha, \gamma) = \int_{-\infty} ^{+\infty} \int_{0}
^{+\infty} \widetilde{g}(x, y, z, t) e^{-i \alpha x -i \gamma z}
dx dz
\end{equation}

\noindent indicates the Laplace-Fourier transform of a general
dependent variable in the $\alpha - \gamma$ phase space and in the
remaining independent variables $y$ and $t$. The governing partial
differential equations are

\begin{eqnarray} \label{IVP2_fou1}
\frac{\partial^2 \hat{v}}{\partial y^2} &-& (k^2 - \alpha_i ^2 + 2
i k cos(\phi) \alpha_i) \hat{v}= \hat{\Gamma} \\
 \nonumber
\frac{\partial \hat{\Gamma}}{\partial t} = &-& (i k cos(\phi) -
\alpha_i) U \hat{\Gamma} + (i k cos(\phi) - \alpha_i) \frac{d^2 U}{dy^2} \hat{v} \\
&+& \frac{1}{Re} [\frac{\partial^2 \hat{\Gamma}}{\partial y^2} -
(k^2 - \alpha_i ^2 + 2 i k cos(\phi) \alpha_i)
\hat{\Gamma}]\label{IVP2_fou2} \\
 \nonumber
 \frac{\partial \hat{\omega}_{y}}{\partial t} = &-& (i k cos(\phi) - \alpha_i) U \hat{\omega}_{y}
  - i k sin(\phi) \frac{dU}{dy} \hat{v} \\
 &+& \frac{1}{Re} [\frac{\partial^2 \hat{\omega}_y}{\partial
y^2} - (k^2 - \alpha_i ^2 + 2 i k cos(\phi) \alpha_i)
\hat{\omega}_y]\label{IVP2_fou3}
\end{eqnarray}

\noindent where $\phi = tan^{-1}(\gamma/\alpha_r)$ is the angle of
obliquity with respect to the $x$-$y$ physical plane, $k =
\sqrt{\alpha_r^2 + \gamma^2}$ is the polar wavenumber and
$\alpha_r = k cos(\phi)$, $\gamma = k sin(\phi)$ are the
wavenumbers in $x$ and $z$ directions respectively. The imaginary
part $\alpha_i$ of the complex longitudinal wavenumber represents
the spatial damping rate in the streamwise direction. In figure
\ref{DT_perturbation_scheme} the three-dimensional perturbative
geometry scheme is depicted.

\noindent From equations (\ref{IVP2_fou1})-(\ref{IVP2_fou3}), it
can be noted that there can neither be advection nor production of
vorticity in the lateral free stream. The vorticity can only be
diffused since only the diffusive terms remains in the limit when
$y \rightarrow \infty$. Perturbation vorticity vanishes in the
free stream, regardless if it is initially inserted there (if
inserted, vorticity is finally dissipated in time when $y
\rightarrow \infty$). This means that the velocity field is
harmonic as $y \rightarrow \infty$.

Governing equations (\ref{IVP2_fou1}), (\ref{IVP2_fou2}) and
(\ref{IVP2_fou3}) need proper initial and boundary conditions to
be solved. Among all solutions, those whose perturbation velocity
field is bounded in the free stream are sought. Periodic initial
conditions for
\begin{equation}
\hat{\Gamma} = \frac{\partial^2 \hat{v}}{\partial y^2} - (k^2 -
\alpha_i ^2 + 2 i k cos(\phi) \alpha_i) \hat{v}
\end{equation}
\noindent can be cast in terms of a set of functions in the $L^2$
Hilbert space, as

$$\hat{v}(0,y) = e^{-(y - y_0)^2} \textmd{cos}(n_0 (y - y_0)) \,\,\,\, \textmd{or}
\,\,\,\, \hat{v}(0,y) = e^{-(y - y_0)^2} \textmd{sin}(n_0 (y -
y_0)),$$
for the symmetric and the asymmetric perturbations, respectively.
Parameter $n_0$ is an oscillatory parameter for the shape
function, while $y_0$ is a parameter which controls the
distribution of the perturbation along $y$ (by moving away, or
bringing nearer, the perturbation maxima from the axis of the
wake). The trigonometrical system is a Schauder basis in each
space $L^p[0,1]$, for $1<p<\infty$. More specifically, the system
$(1, sin(n_0 y), cos(n_0y), \dots)$, where $n_0=1,2,\dots$, is the
Schauder basis for the space of square-integrable periodic
functions with period $2\pi$. This means that any element of the
space $L^2$, where the dependent variables are defined, can be
written as an infinite linear combination of the elements of the
basis.

\noindent The transversal vorticity $\hat{\omega}_y$ is chosen
initially equal to zero throughout the $y$ domain in order to
ascertain which is the net contribution of three-dimensionality on
the transversal vorticity generation and temporal evolution.
However, it can be demonstrated that the eventual introduction of
an initial transversal vorticity does not actually affect the
perturbation temporal evolution.

Once initial and boundary conditions are properly set, the partial
differential equations (\ref{IVP2_fou1})-(\ref{IVP2_fou3}) are
numerically solved by the method of lines. The spatial derivatives
are centre differenced and the resulting system is then integrated
in time by an adaptative multi-step method (variable order
Adams-Bashforth-Moulton PECE solver). The transversal
computational domain is large thirty body scales. By enlarging the
computational domain to 50 and 100 body scales the results vary on
the third  and fourth significant digit, respectively.

\subsection{Measure of the growth}\label{Measure_Growth}

One of the salient aspects of the initial-value problem is to
observe the early transient evolution of various initial
conditions. To this end, a measure of the perturbation growth can
be defined through the disturbance kinetic energy density in the plane $(\alpha, \gamma)$ (see e.g \cite{SH01},\cite{CJJ03})

\begin{eqnarray}
\nonumber e(t; \alpha, \gamma, Re) &=& \frac{1}{2} \frac{1}{2y_d}
\int_{-y_d}^{+y_d} (|\hat{u}|^2
+ |\hat{v}|^2 + |\hat{w}|^2) dy \\
&=& \frac{1}{2} \frac{1}{2y_d} \frac{1}{|\alpha^2 +
\gamma^2|}\int_{-y_d}^{+y_d} (|\frac{\partial \hat{v}}{\partial
y}|^2 + |\alpha^2 + \gamma^2| |\hat{v}|^2 + |\hat{\omega}_y |^2)
dy,
\end{eqnarray}

\noindent where $2y_d$ is the extension of the spatial numerical
domain. The value $y_d$ is defined so that the numerical solutions
are insensitive to further extensions of the computational domain
size. Here, we take $y_d = 15$. The total kinetic energy can be
obtained by integrating the energy density over all $\alpha_r$ and
$\gamma$. The  amplification factor $G(t)$ can be
introduced in terms of the normalized energy density

\begin{equation}
G(t; \alpha, \gamma) = \frac{e(t; \alpha, \gamma)}{e(t=0; \alpha,
\gamma)}.
\end{equation}

\noindent This quantity can effectively measure the growth of a
disturbance of wavenumbers $(\alpha, \gamma)$ at the time $t$, for
a given initial condition at $t = 0$ (Criminale \textit{et al.}
1997; Lasseigne \textit{et al.} 1999).

The temporal growth rate on the kinetic energy $r$

\begin{equation}
r(t; \alpha, \gamma) = \frac{log|e(t; \alpha, \gamma)|}{2t},
\;\;\; t>0 \label{IVP2_tgr}
\end{equation}

\noindent is introduced in order to evaluate both the early
transient as well as the asymptotic behavior of the
perturbations (here it is  the first moment of the perturbation which is assumed to asymptotically approach an exponential growth). Computations to evaluate the long time asymptotics
are made by integrating the equations forward in time beyond the
transient (\cite{CJLJ97}, \cite{LJJC99}) until the temporal growth
rate $r$ asymptotes to a constant value ($dr/dt < \epsilon$, where
$\epsilon$ is of the order $10^{-4}$). The angular frequency
(pulsation) $\omega$ of the perturbation can be introduced by
defining a local, in space and time, time phase $\varphi$ of the
complex wave at a fixed transversal station (for example $y = 1$)
as
\begin{eqnarray}
\hat{v}(y, t; \alpha, \gamma, Re) = A_t(y; \alpha, \gamma, Re) e^{i \varphi(t)}
\end{eqnarray}

\noindent and then computing the time derivative of the phase
perturbation $\varphi$ (\cite{W74})

\begin{eqnarray}
\omega(t) = \frac{d \varphi(t)}{dt}.
\end{eqnarray}

\noindent Since $\varphi$ is defined as the phase variation in
time of the perturbative wave, it is reasonable to expect constant
values of frequency, once the asymptotic state is reached.

\section{Results}\label{Results}

We present a summary of the most significant transient behavior
and asymptotic fate of the three-dimensional perturbations. The
temporal evolution is observed in the intermediate asymptotic
region of the wake, which is the region where the spatial
evolution is predominant. It can be demonstrated that changing the
number of oscillations $n_0$ and the parameter $y_0$ that controls
the perturbation distribution along the $y$ direction can only
extend or shorten the duration of the transient, while the
ultimate state is not altered. More specifically, if the
perturbation oscillates rapidly or is concentrated mainly outside
the shear region of the basic flow, for a stable configuration,
the final damping is accelerated while, for an unstable
configuration, the asymptotic growth is delayed. Thus, these two
parameters are not crucial, because their influence can be
recognized a priori. Therefore, in the following we use the two
reference values, $n_0 = 1$ and $y_0 = 0$, and focus the attention
mainly on parameters such as the obliquity, the symmetry, the
value of the polar wavenumber and the spatial damping rate of the
disturbance. In particular, the polar wavenumber $k$ changes in a
range of values reaching at maximum the order of magnitude $O(1)$,
according to what is suggested by recent modal analyses
(\cite{TSB06}, \cite{BT06}). The order of magnitude of the spatial
damping rate $\alpha_i$ varies around the polar wavenumber value.

\subsection{Exploratory analysis of the transient dynamics}\label{Transient}

\begin{figure}[!h]
\begin{minipage}[]{0.5\columnwidth}
   \includegraphics[width=\columnwidth]{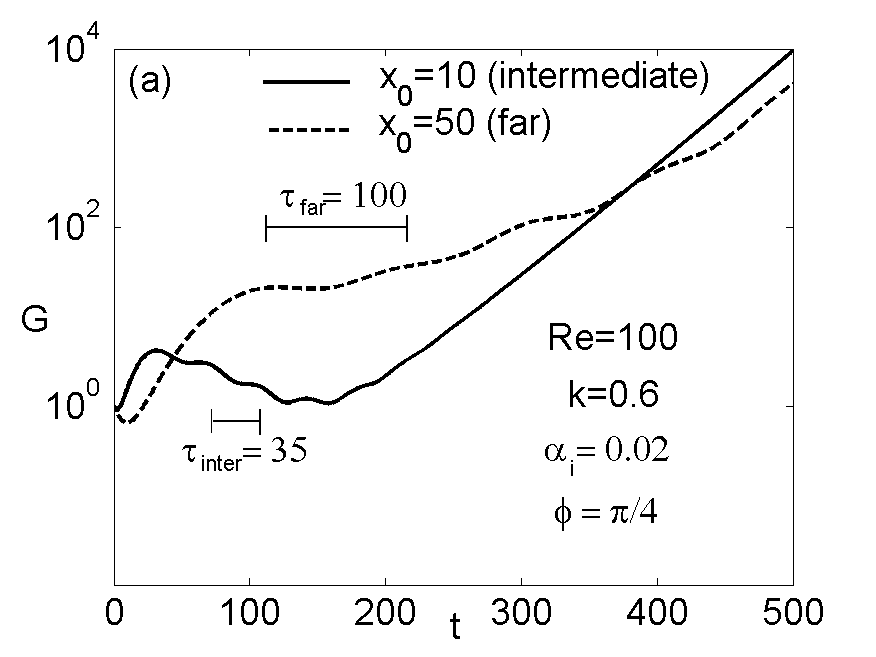}
    \label{DT_asym_pi_4}
\end{minipage}
\begin{minipage}[]{0.5\columnwidth}
   \includegraphics[width=\columnwidth]{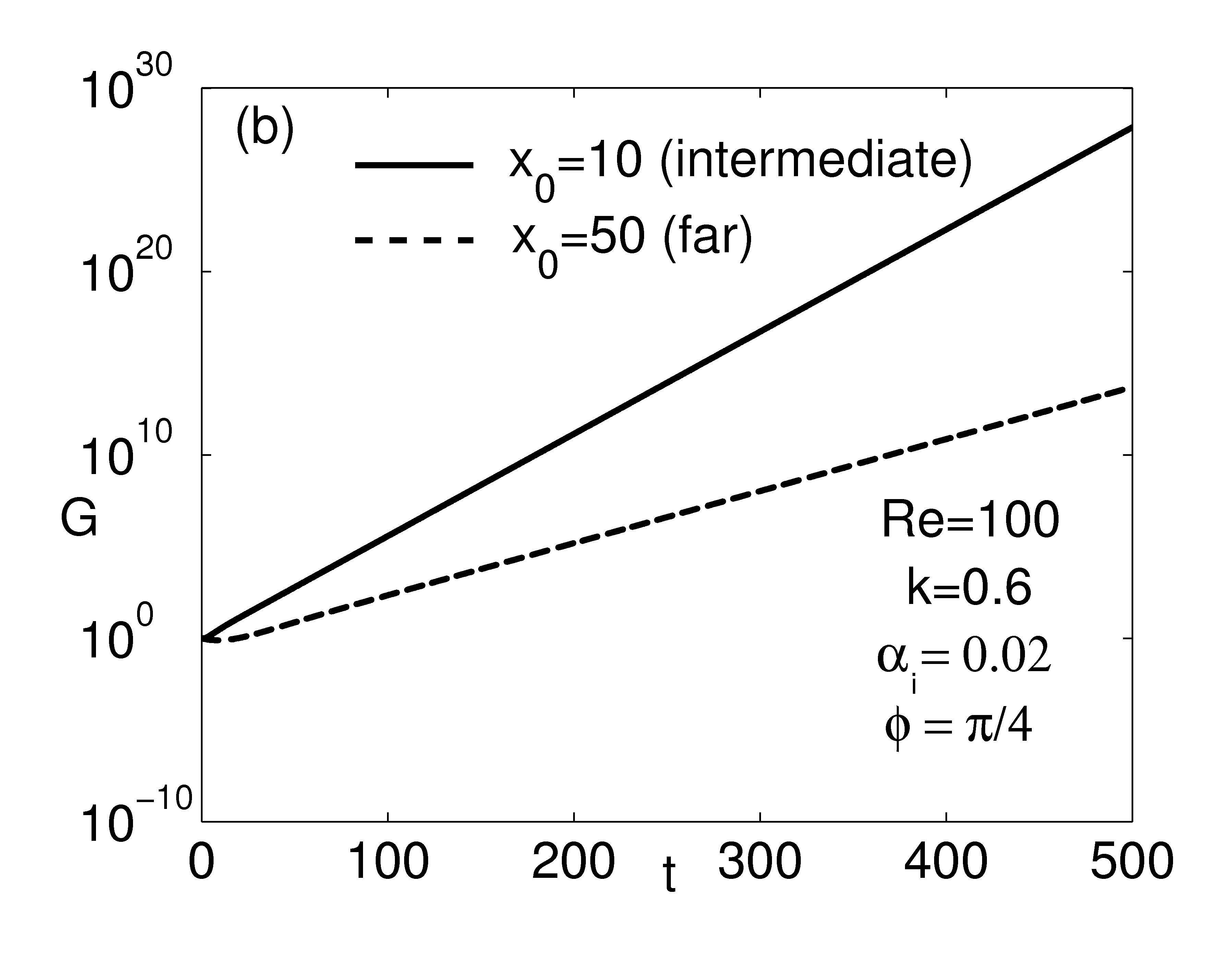}
    \label{DT_sym_pi_4}
\end{minipage}
\begin{minipage}[]{0.5\columnwidth}
   \includegraphics[width=\columnwidth]{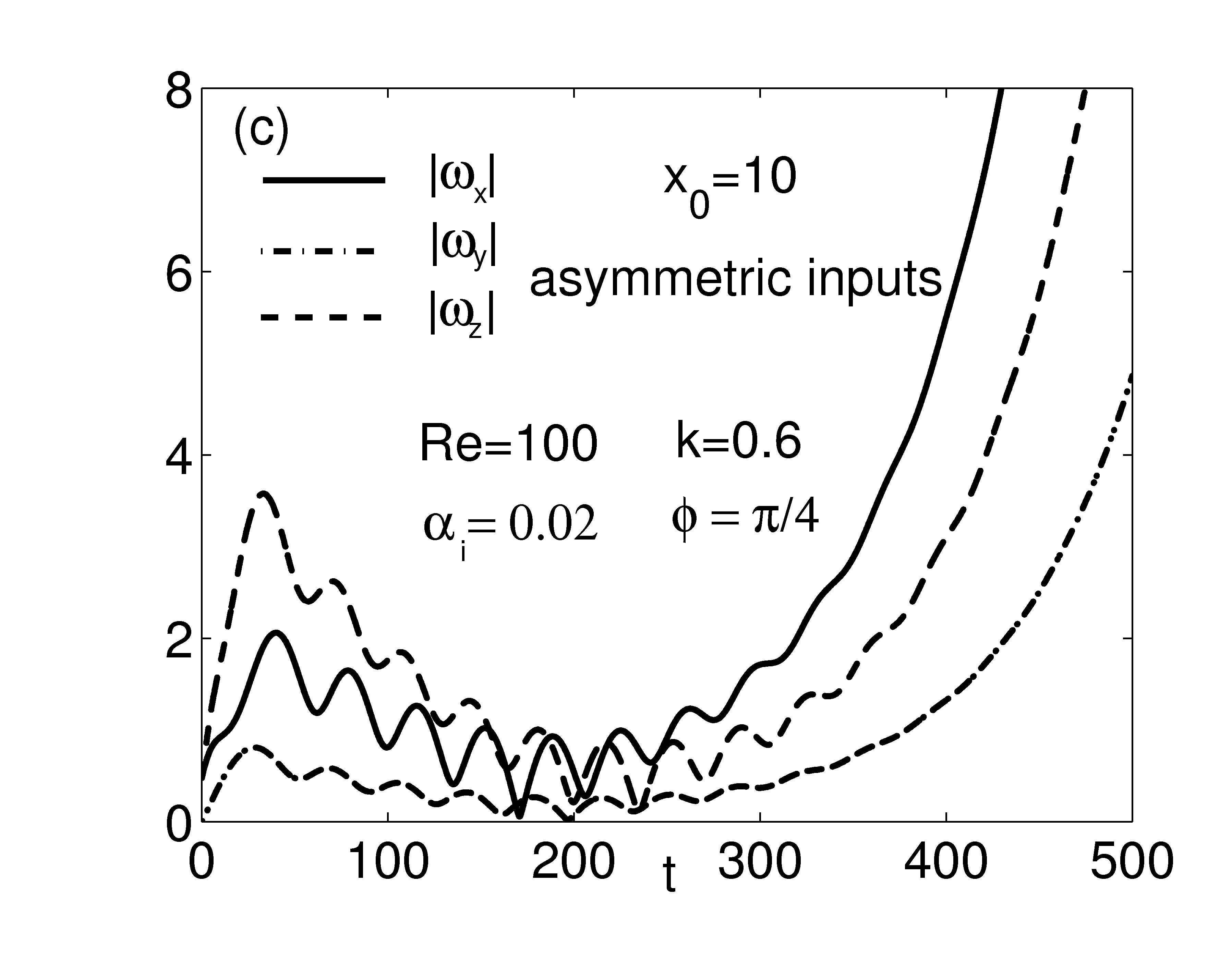}
    \label{DT_asym_pi_4}
\end{minipage}
\begin{minipage}[]{0.5\columnwidth}
   \includegraphics[width=\columnwidth]{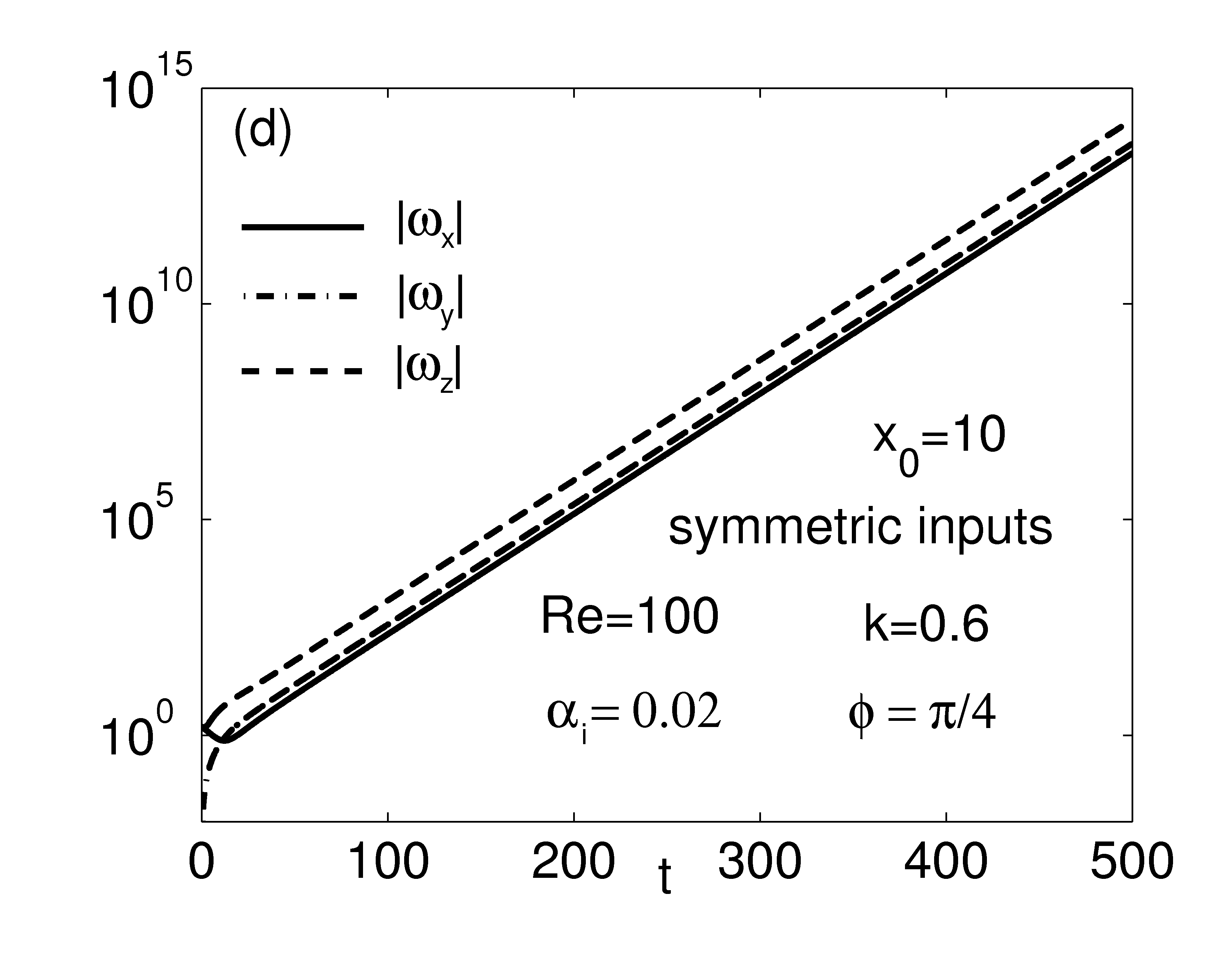}
    \label{DT_sym_pi_4}
\end{minipage}\caption{Effect of the symmetry of the perturbation. (a) - (b): The
amplification factor $G$, and (c)-(d) the perturbation vorticity
components at $y = 1$ as a function of time. (a)-(c) asymmetric
initial condition, (b)-(d) symmetric initial condition.
Intermediate ($x_0=10$, solid curves) and far field ($x_0=50$,
dashed curves) wake configurations. The periods $\tau_{inter},
\tau_{far}$ are the periods of the modulation visible on $G$, in
the intermediate and far field, respectively.
The values of the vorticity component in
part (d) have no physical meaning. The plot simply shows that, on
the contrary of the asymmetric case in part(c), the symmetric
disturbance growth has a short transient after which it becomes
homogeneous in time.} \label{DT_asym_sym}
\end{figure}

Figure \ref{DT_asym_sym} takes into account the influence, on the
early time behavior, of the perturbation symmetry and of the wake
region considered in the analysis, which is represented by the
parameter $x_0$. All the configurations considered are
asymptotically amplified, but the transients are different.
The asymmetric cases (a) present, for both the intermediate
position $x_0 = 10$ (solid curve) and  the far field position $x_0
= 50$ (dashed curve), two temporal evolutions. For $x_0=10$ a
local  maximum, followed by a minimum, is visible in the energy
density, then the perturbation is slowly amplifying and the
transient can be considered extinguished only after hundreds of
time scales. For $x_0=50$ these features are less marked. It can
be noted that the far field configuration ($x_0 = 50$) has a
faster growth than the intermediate field configuration ($x_0 =
10$) up to $t = 400$. 
Beyond this instant
the growth related to the intermediate configuration will prevail
on that of the far field configuration.
In the symmetric cases
(b) the growths become monotone after few time scales ($t = 20$)
and the perturbations quickly reach their asymptotic states
(around $t = 50$). The intermediate field configuration ($x_0 =
10$, solid curve) is always growing faster than the far field
configuration ($x_0 = 50$, dashed curve). This particular case
shows a behavior that is generally observed in this analysis,
that is, asymmetric conditions lead to transient evolutions that
last longer than the corresponding symmetric ones, and
demonstrates that the transient growth for a longitudinal station
in the far wake can be faster than in the intermediate wake.
It should be noted that, even if the
asymmetric perturbation leads to a much slower  transient
growth than that observed for the symmetric case, the growth rate become equal when the asymptotic states  are reached (see
for example fig. \ref{IVP_comparison_NMT_IVP_kfixed}). The
temporal window  shown  in fig.3 ($t=500$) does not yet capture the
asymptotic state of the asymmetric input. However, we
observed that further in time the amplification factor
$G$ reaches the same  order of magnitude  of the
symmetric perturbation.

\begin{figure}[!b]
   \includegraphics{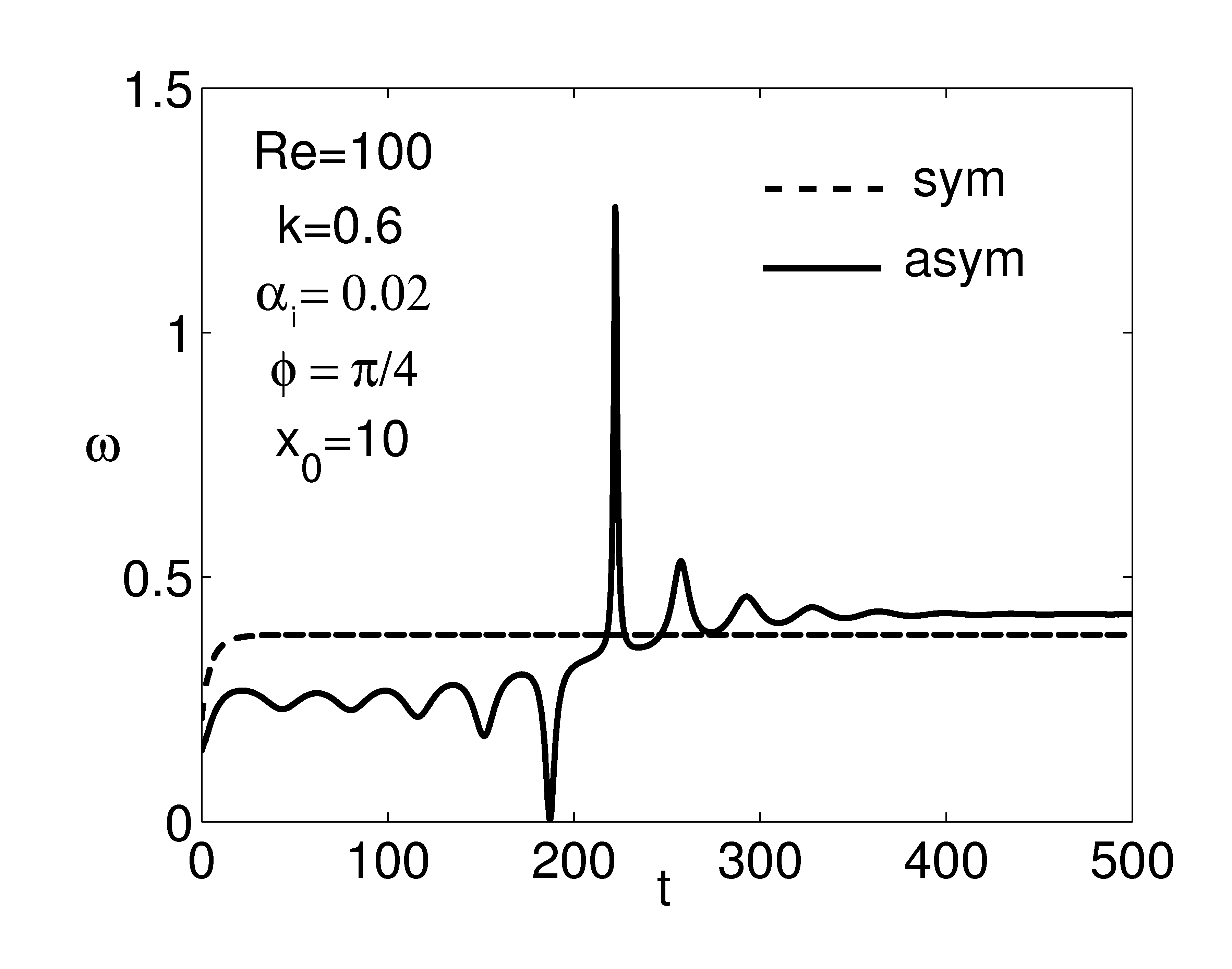}
\caption{Pulsation behavior. The wave parameters are those shown
in the previous figure, for $x_0=10$.} \label{pulsation}
\end{figure}

The more noticeable results presented in fig. 3 are that the
asymmetric growths in the early transient are much less rapid than
the symmetric ones and that the function $G$, in the case of
asymmetric perturbations only, shows a modulation which is very
evident in the first part of the transient, and which corresponds
to a modulation in amplitude of the pulsation of the instability
wave, see fig.\ref{pulsation}. In fact, the pulsation varies: in
the early transient it oscillates around a mean value with a
regular period, which is the same visible on G, and with an amplitude which is growing until
this value jumps to a new value around which oscillates in a
damped way. This second value is the asymptotic constant value.
This behavior is always observed in the case of asymmetric
longitudinal or oblique instability waves. Instead, it is not
shown by transversal ($\phi = \pi/2$) waves or by symmetric waves,
see fig. 4, where, on the one hand, the asymptotic value, nearly
equal to that of the asymmetric perturbation,  is rapidly reached
after a short monotone growth and where, on the other, the growth
is many orders of magnitude faster, and as a consequence, a
modulation would not be easily observable. Thus, we may comment
here on the fact that two time scales are observed in the
transient and long term behavior of longitudinal and oblique
perturbations: namely, the periodicity associated to the average
value of the pulsation in the early transient, clearly visible in
the asymmetric case only, and the final asymptotic pulsation. The
asymptotic value of the pulsation is higher than the initial one,
typically  is about $2.5$ times higher. The period of the
frequency modulation of the energy density $G$ is larger, nearly
(1.4 - 1.7) higher, than the  average period of the oscillating
wave in the early transient, because $G$ is a square norm of the
system solution. Thus the evolution of the system exhibits two
periodic patterns at different frequencies: the first, of
transient nature, and the other of asymptotic nature. When the
average damping of the energy density in the early transient is
not strong and is then followed by a monotone asymptotic growth,
the  change of the frequency of the oscillation is evident, see
fig.\ref{pulsation}.

This kind of behavior is often observed in the study of linear
systems with an oscillating norm, a problem that naturally arises
in the context of the linearized formulation of
convection-dominated systems over finite length domains. The
occurrence of oscillating patterns in the energy evolution of the
solutions is linked to the non-normal character of the linear
operator which describes the system, see e.g. Coppola and De Luca
(2006)\cite{CDL06}.

\begin{figure}[!b]
\begin{minipage}[]{0.5\columnwidth}
   \includegraphics[width=\columnwidth]{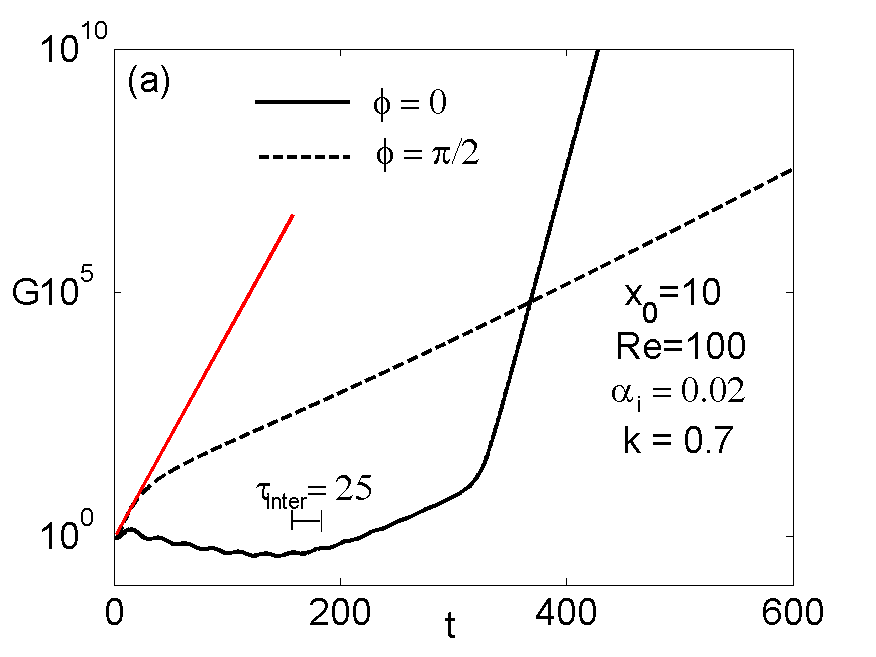}
    \label{DT_G_2D3D}
\end{minipage}
\begin{minipage}[]{0.5\columnwidth}
   \includegraphics[width=\columnwidth]{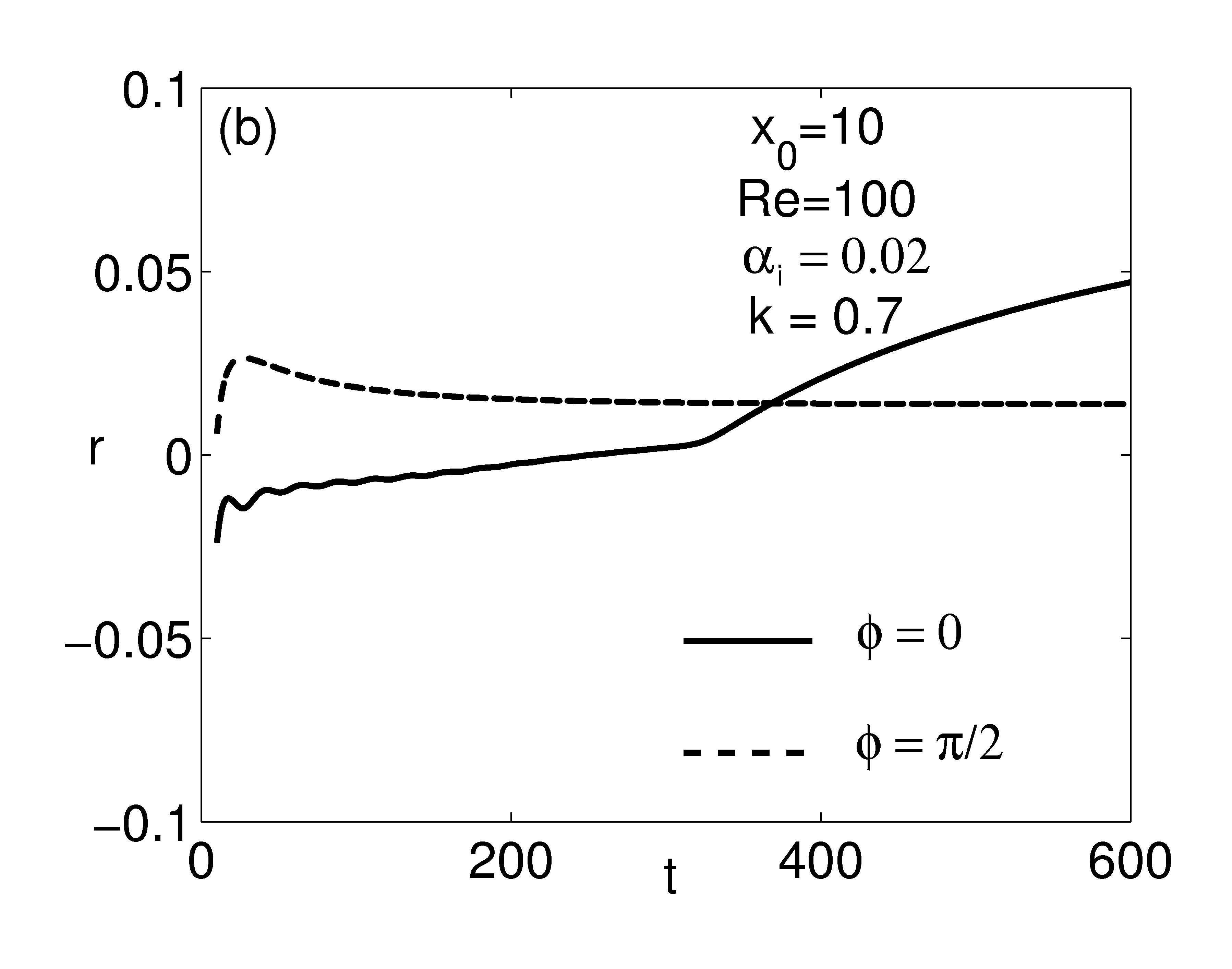}
    \label{DT_r_2D3D}
\end{minipage}
\begin{minipage}[]{0.5\columnwidth}
   \includegraphics[width=\columnwidth]{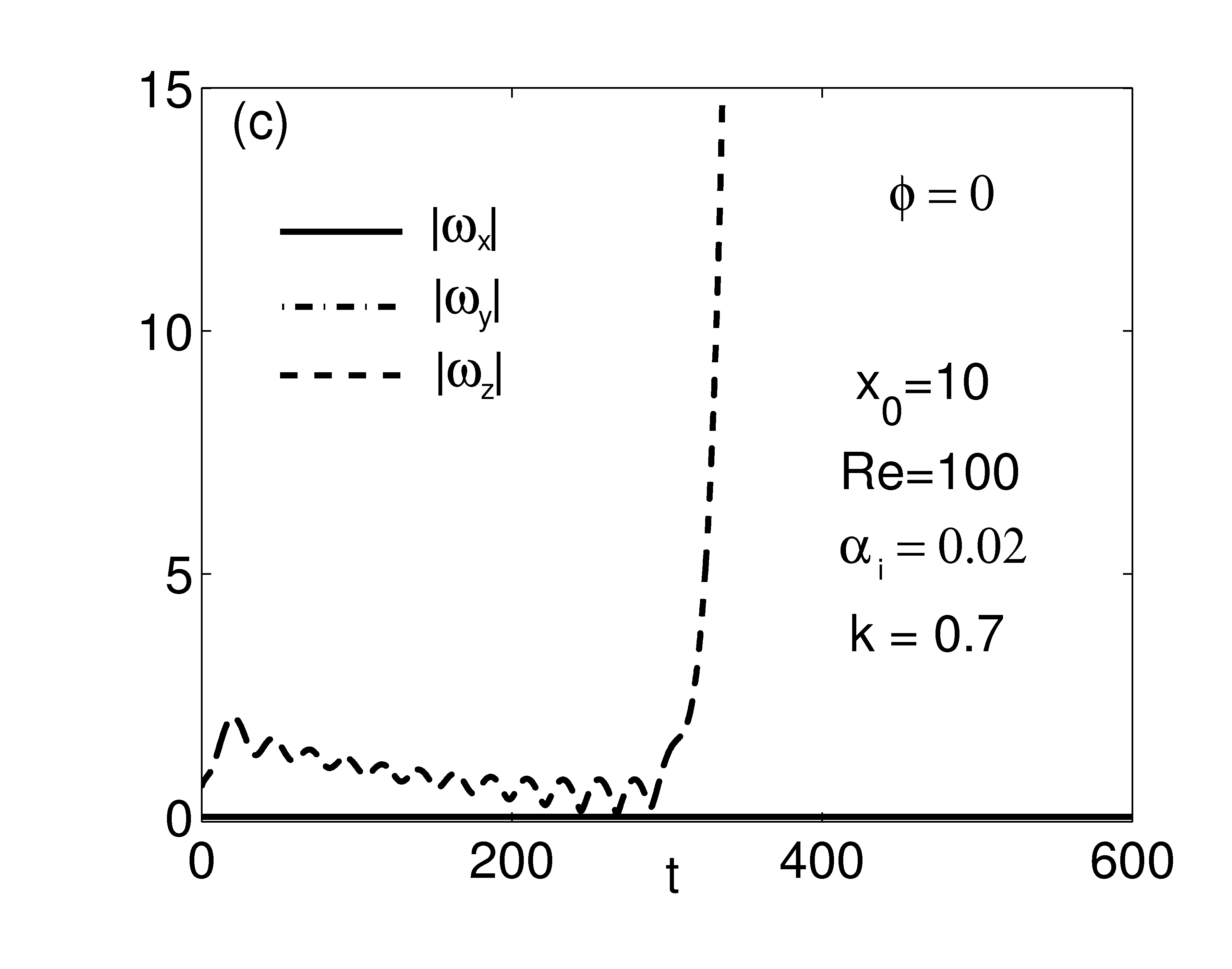}
    \label{DT_G_2D3D}
\end{minipage}
\begin{minipage}[]{0.5\columnwidth}
   \includegraphics[width=\columnwidth]{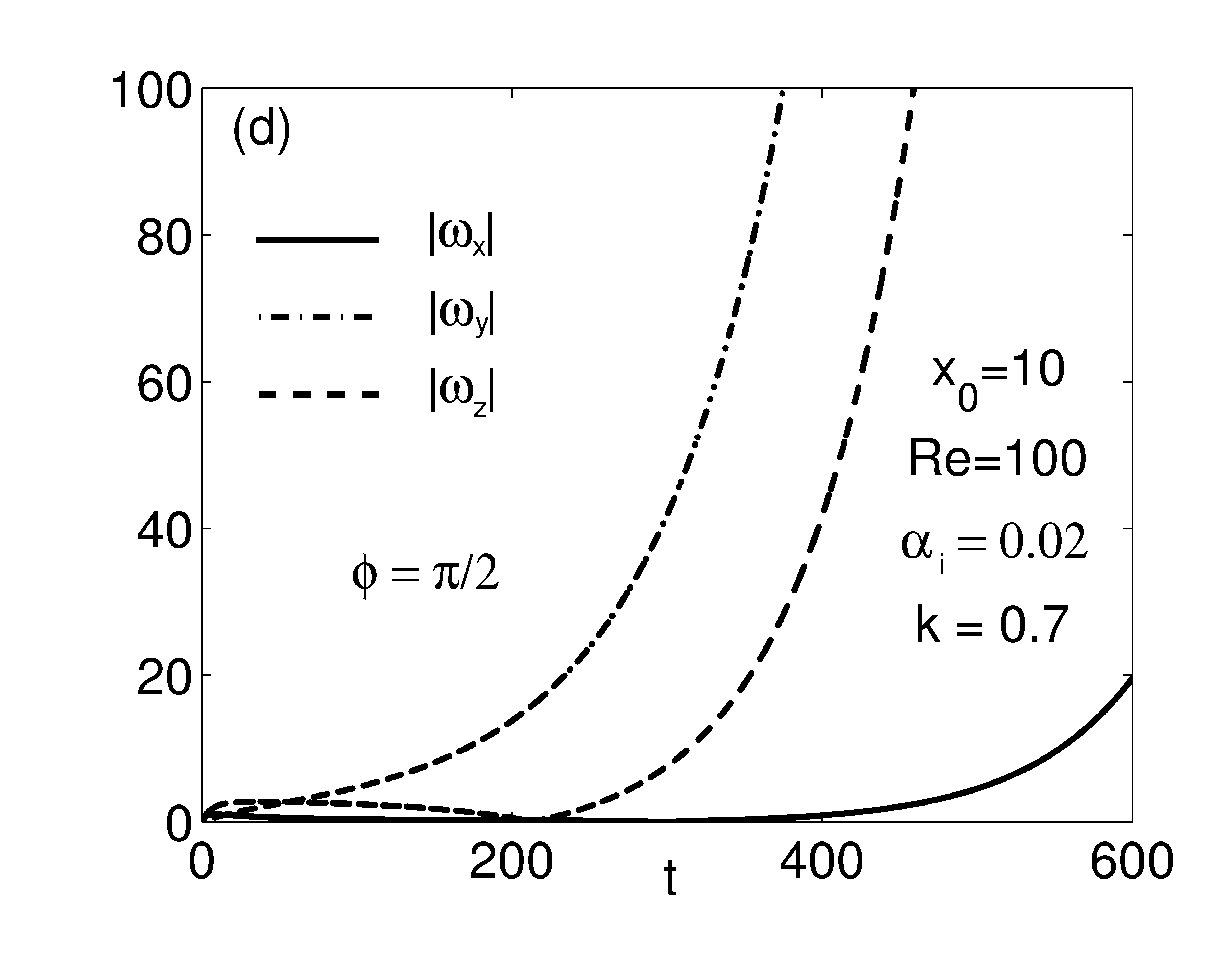}
    \label{DT_r_2D3D}
\end{minipage}
\caption{Effect of the angle of obliquity $\phi$. (a) The
amplification factor $G$ and (b) the temporal growth rate $r$ as
functions of time. Asymmetric initial condition, $\phi = 0$ (solid
curves), $\phi = \pi/2$ (dashed curves). (c)-(d) the perturbation
vorticity components as functions of time at $y = 1$, (c) $\phi=0$
and (d) $\phi=\pi/2$. The periods $\tau_{inter}$ is the period of
the modulation visible on $G$, in the intermediate field.
}\label{DT_G_r_2D3D}
\end{figure}

Figure \ref{DT_G_r_2D3D} illustrates an  interesting comparison
between two-dimensional and three-dimensional waves (note that a
logarithmic scale on the ordinate is used in part (a) of Fig.
\ref{DT_G_r_2D3D}). The purely two-dimensional wave (solid curve)
is rapidly reaching a first maximum of amplitude (at about $t =
15$), then the perturbation decreases while oscillating and
reaches a minimum around $t = 150$. Afterwards, the disturbance
slowly grows up to $t \approx 300$, where an inflection point of
the amplification factor $G$ occurs. It should be noted that this
behavior is controlled solely by the evolution of the $\omega_z$,
that is by the vorticity component present in the basic flow only.
Then, the growth becomes faster and the perturbation is highly
amplified in time. The purely orthogonal perturbation (dashed
curve) is instead immediately amplified. The trend is monotone,
and does not present visible fluctuations in time. The initial
growth is actually rapid and an inflection point of the
amplification factor $G$ can be found around $t = 50$. Beyond this
point, the growth changes its velocity and becomes slower, but
still destabilizing. Both cases have asymmetric initial conditions
and are ultimately amplified. In agreement with Squire theorem,
the two-dimensional case turns out to be more unstable than the
three-dimensional one, as the 2D asymptotically established
exponential growth is more rapid than the 3D one (see solid and
dashed curves in Fig. \ref{DT_G_r_2D3D}(a) for $t
> 400$). However, it should be noted that, for an extended
part of the transient (up to about $t \approx 380$), the
three-dimensional perturbation presents a larger growth than the
two-dimensional one. 

\begin{figure}[!h]
\begin{minipage}[]{0.5\columnwidth}
   \includegraphics[width=\columnwidth]{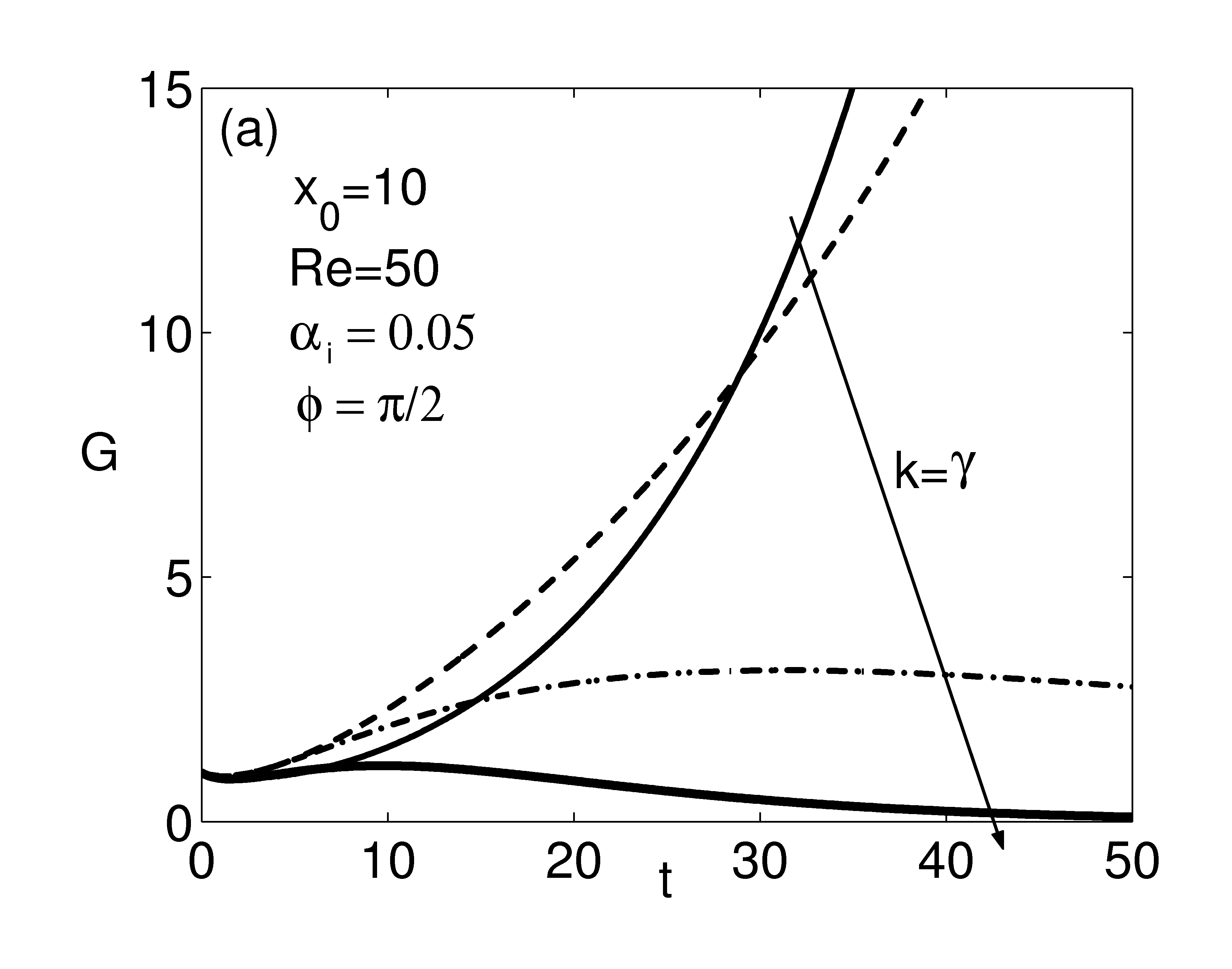}
    \label{G_k}
\end{minipage}
\begin{minipage}[]{0.5\columnwidth}
   \includegraphics[width=\columnwidth]{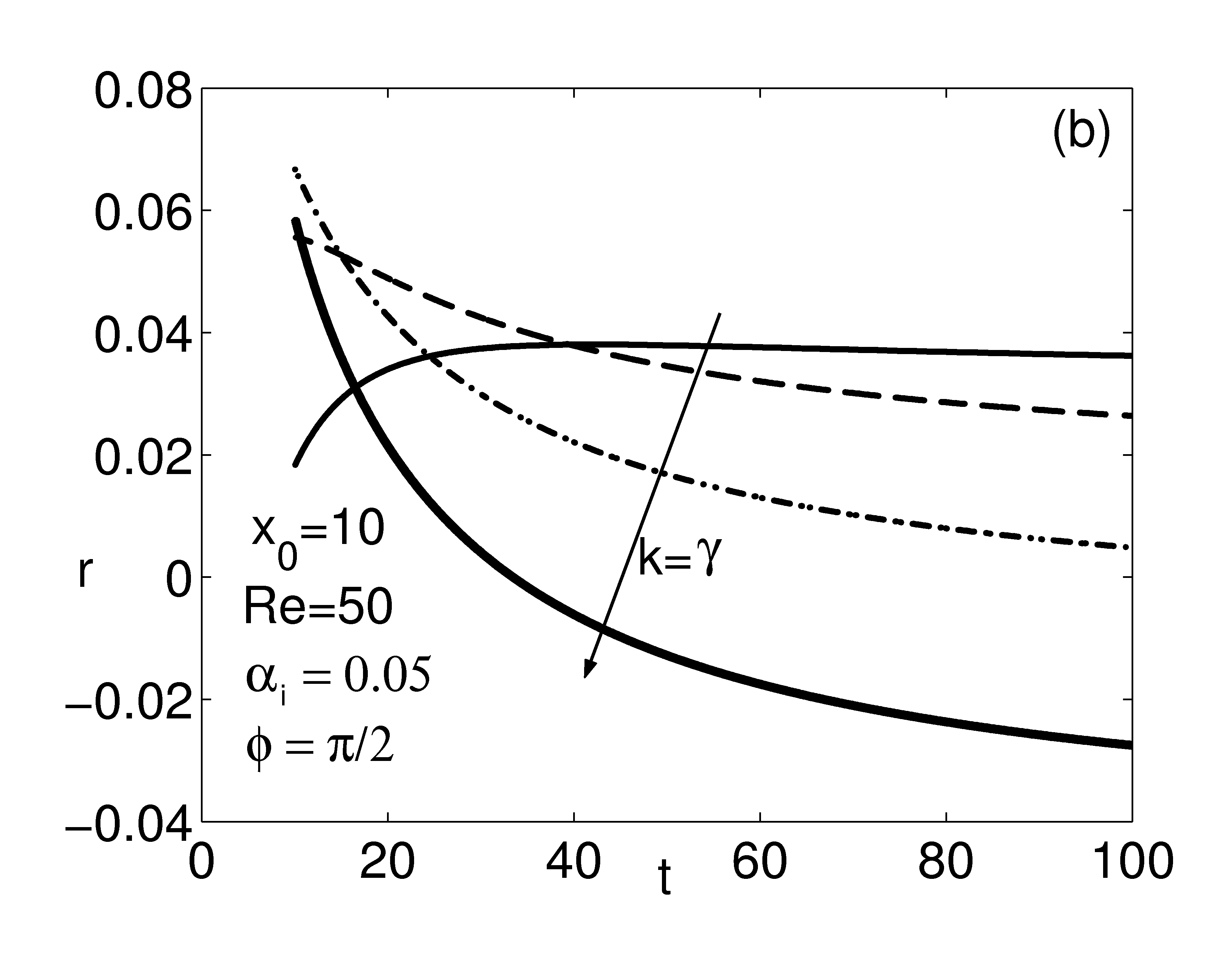}
    \label{r_k}
\end{minipage}
\caption{Effect of the polar wavenumber $k$. (a) The amplification
factor $G$ and (b) the temporal growth rate $r$ as function of
time. Symmetric initial condition, $k = 0.5, 1, 1.5, 2$.}
\label{G_r_k}
\end{figure}

\begin{figure}[!h]
\begin{minipage}[]{0.5\columnwidth}
   \includegraphics[width=\columnwidth]{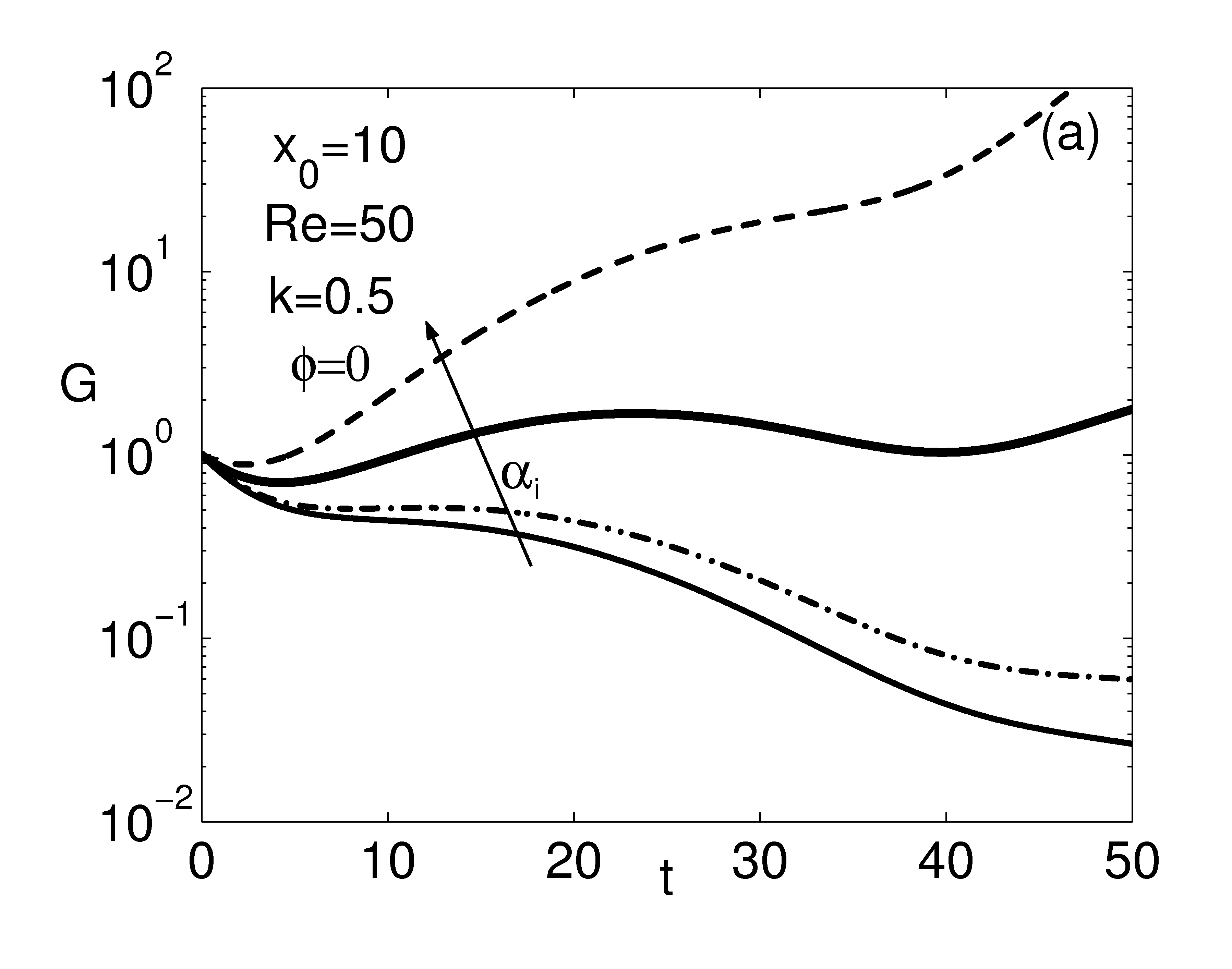}
    \label{G_k}
\end{minipage}
\begin{minipage}[]{0.5\columnwidth}
   \includegraphics[width=\columnwidth]{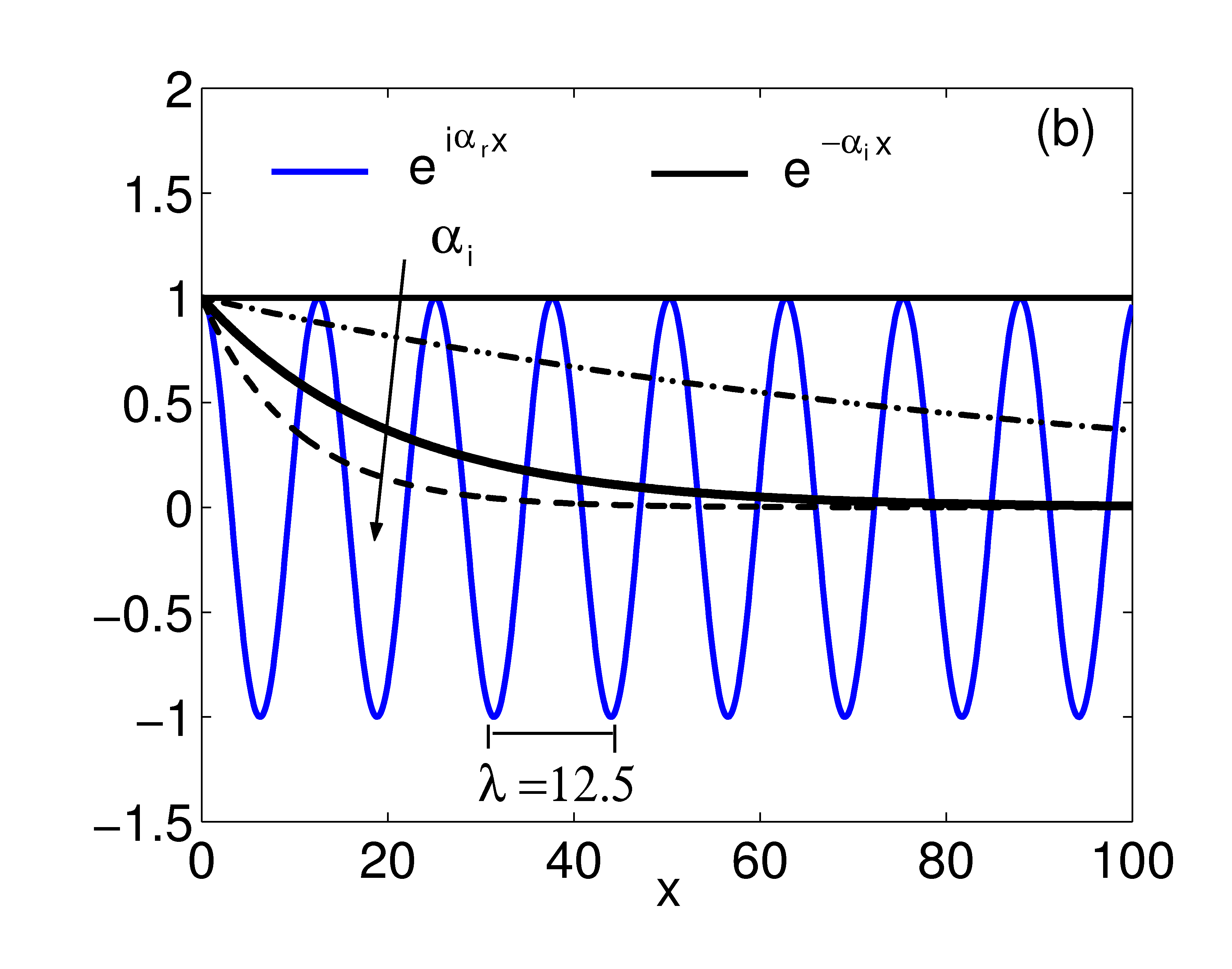}
    \label{r_k}
\end{minipage}
\caption{Effect of the spatial damping rate $\alpha_i$. (a) The
amplification factor $G$ as function of time and (b) the wave
spatial evolution in the $x$ direction for $k=\alpha_r=0.5$.
Asymmetric initial condition, $\alpha_i = 0, 0.01, 0.05, 0.1$.}
\label{G_r_alphai}
\end{figure}

Fig. \ref{G_r_k} demonstrates that purely orthogonal
three-dimensional unstable perturbations may become damped by
increasing their wavenumber ($k=\gamma$). In the case displayed in
this figure, this happens when the wave number is increased beyond
the value $1$. Before the asymptotic stable states are reached,
these configurations yield maxima of the energy density (e.g. when
$k=1.5, G \sim 3$ at $t\sim 30$) in the transients. This trend is
also typical of oblique and longitudinal waves, and it can be
considered a universal feature in the context of the stability of
near parallel shear flows. It should be noted that, in fig.
\ref{G_r_k}, the perturbation is symmetric and again the amplitude
modulation is not observed in the early transient, even in the
asymptotically stable situations. However, this example of
transient behavior also contains a feature which is specific of
orthogonal, both amplified or damped, and symmetric or asymmetric,
perturbations, namely, the fact the most amplified component of
the vorticity is the $\omega_y$, see also fig.\ref{DT_G_r_2D3D},
part d.

In fig. \ref{G_r_alphai} a significant phenomenon is observed for
a longitudinal wave. 
By changing the
order of magnitude of $\alpha_i$, it can be seen that perturbations that are more
rapidly damped in space (see, in fig. \ref{G_r_alphai}b, the longitudinal spatial evolution of the wave) yield  a faster growth in
time. In fact, for nearly uniform waves in $x$ direction
($\alpha_i\rightarrow0$) the configurations are asymptotically
damped in time, while for increasing values of the spatial damping
rate the perturbations are amplified in time (note that a
logarithmic scale is used on part (a) of the figure).  A possible general explanation is that the introduction in a physical system of a spatial concentration  of kinetic energy is always destabilizing, hence the higher is the disturbance concentration the faster is  the growth factor.

\subsection{Physical interpretation of the different growth rate of symmetric and asymmetric disturbances}\label{Physics}

The dramatically increased growth rate of the symmetric mode with respect to the asymmetric mode can be understood from the induced velocity of the vorticity field.  For simplicity, imagine that the wake consists of two parallel shear layers of opposite vorticity, as shown in figure 8(a).  Further assume that the vorticity is discretized into a finite number of identical vortices.  Suppose the upper shear layer is perturbed into a sinusoidal shape, as shown in figure 8(b).  The induced velocity at the crest of the sinusoid at point 1 due to the other vortices in the upper shear layer alone is up and to the right as indicated by the arrow, corresponding to the classical Kelvin-Helmholtz instability.

Now consider the asymmetric mode, so that the lower shear layer is a sinusoid in phase with the upper one (figure 8(c)).  The induced velocity at the crest of the lower shear layer at point 2 due to the other vortices in the lower shear layer alone is exactly the opposite of that at point 1.  Thus the growth in the asymmetric mode is due to only higher order gradients in the induced velocity field of one shear layer on the other.

In contrast, with a symmetric perturbation, the lower shear is the mirror image of the upper.  In figure 8(d), the induced velocity at point 2 is necessarily the mirror image of that at point 1.  Both points move in concert to the right, without the low order velocity cancellation of the asymmetric mode.  Thus the symmetric mode grows much faster.

\begin{figure}[!h]
  \centerline{\includegraphics[height=10cm]{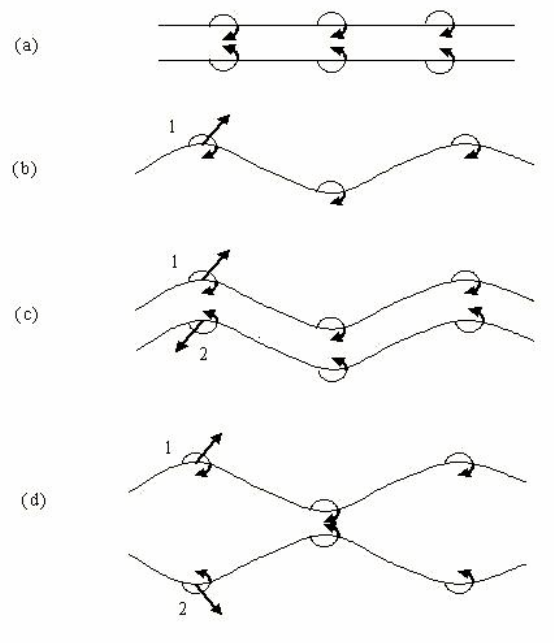}}
  \caption{Induced velocity field of the symmetric and asymmetric modes. (a): Idealization of the base
  flow of the wake into two shear layers of opposite sign. (b): Perturbed upper shear layer. (c):
  Induced velocities in the asymmetric mode. Note that they cancel to lowest order. (d): Induced velocities in
the symmetric mode. Superposition enhances the amplitude at both
points 1 and 2.}\label{vortices}
\end{figure}

This linear absolute instability takes place in the intermediate and far field and acts as a source of excitation for the pair of steady recirculating eddies in the lee of the cylinder.
The onset of a time periodic flow, a supercritical Hopf bifurcation (\cite{J87}, \cite{PMB87}) indicates that in the end the vortices are shed alternatively from the separated streamlines above and below the cylinder forming the Von Karman vortex street. The vortex street is the stable configuration  after the bifurcation has taken place. It has the symmetry of a traveling sinuous mode, which indicates  asymmetry up-and-down of the cylinder. Thus, it can be observed that, also in the context of the vortex shedding, asymmetry  shows higher stability properties with respect to symmetry.

\subsection{Asymptotic fate and comparison with modal analysis}\label{Asymptotic}

\begin{figure}[!h]
\vspace{-0.2cm}
\begin{minipage}[]{0.5\columnwidth}
   \includegraphics[width=\columnwidth]{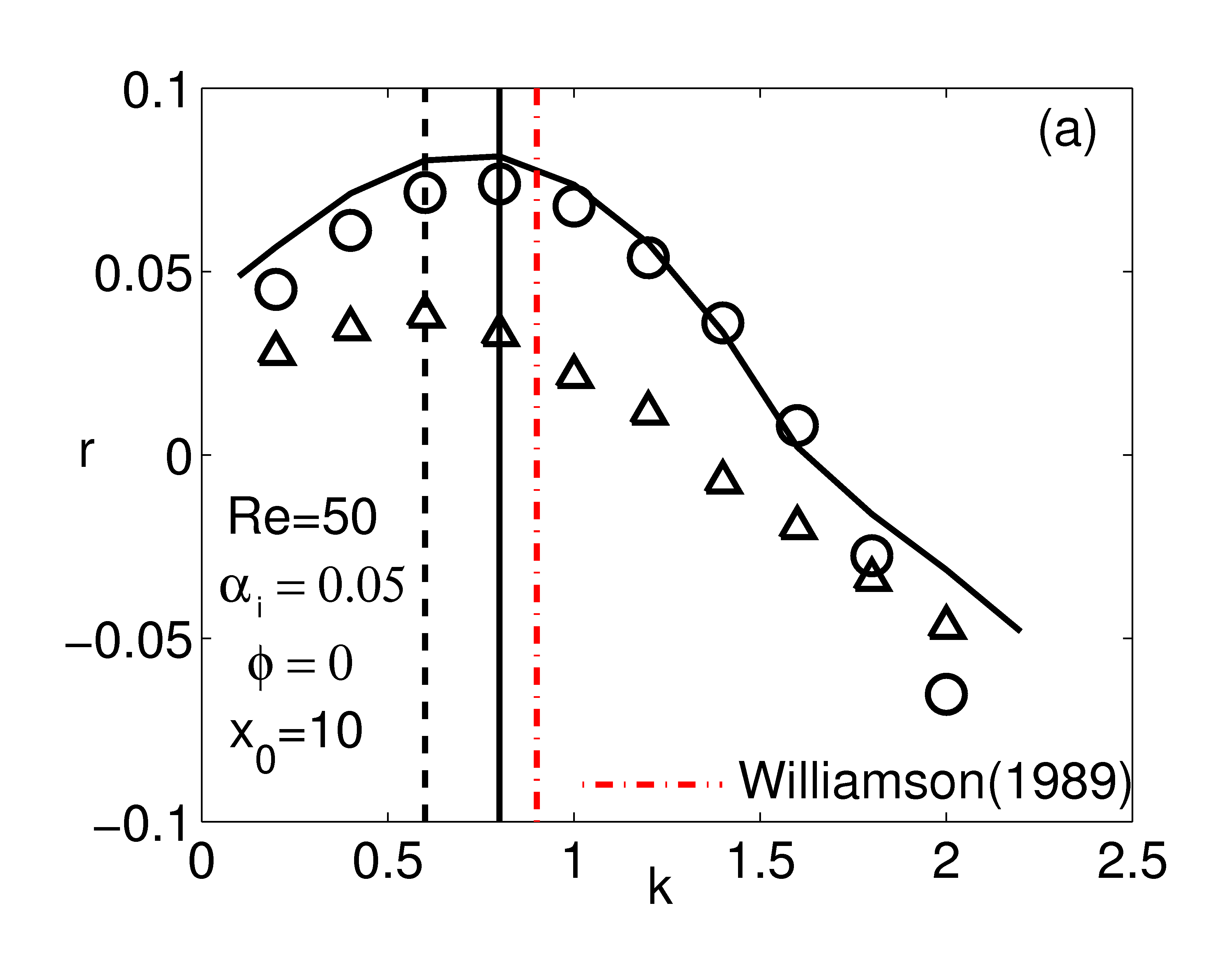}
    \label{DT_r35_full}
\end{minipage}
\vspace{-0.2cm}
\begin{minipage}[]{0.5\columnwidth}
   \includegraphics[width=\columnwidth]{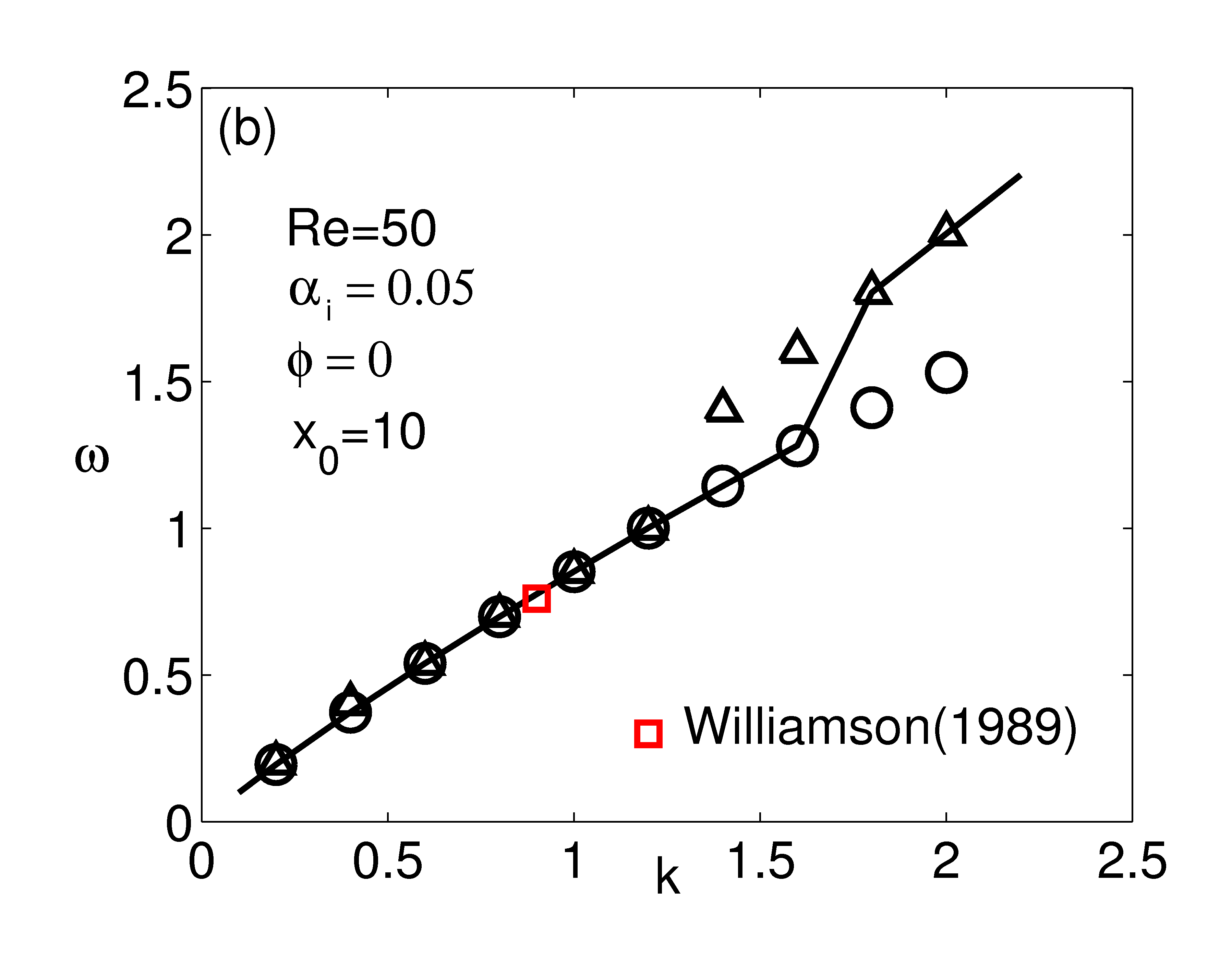}
    \label{DT_om35_full}
\end{minipage}
\caption{$\textsc{(a)}$ Temporal growth rate and $\textsc{(b)}$
pulsation.  Comparison among the asymptotic results obtained by
the IVP analysis (circles: symmetric perturbation; triangles:
asymmetric perturbation) and the normal mode analysis (solid
lines, see \cite{TSB06}. The asymptotic values for the IVP
analysis are determined when the condition $dr/dt < \epsilon$
($\epsilon \sim 10^{-4}$) is satisfied. For the symmetric
perturbations the asymptote was reached before 50 time scales were
elapsed, for the asymmetric perturbation the asymptote was
determined after 500 time scales were elapsed.}
\label{IVP_comparison_NMT_IVP_kfixed}
\end{figure}

Figure \ref{IVP_comparison_NMT_IVP_kfixed} presents a comparison between the initial-value problem and the asymptotic
theory results represented by the zero order Orr-Sommerfeld
problem (\cite{TSB06}) in terms of temporal growth rate $r$ and
pulsation $\omega$.

\noindent In fig. \ref{IVP_comparison_NMT_IVP_kfixed} the
imaginary part $\alpha_i$ of the complex longitudinal wavenumber
is fixed, and differing polar wavenumbers ($k = \alpha_r$) are
considered. For both the symmetric and asymmetric arbitrary
disturbances here considered, a  good agreement with the stability
characteristics given by the multiscale near-parallel
Orr-Sommerfeld theory can be observed. However, it should be noted
that the wave number corresponding to the maximum growth factor in
the case of asymmetric perturbations is about  15\% lower than
that obtained in the case of symmetric perturbations and that
obtained by the normal mode analysis. When the  perturbations are
asymmetric, the transient is very long, of the order of hundreds
time scales. This difference can be due either to the fact that
the true asymptote was not yet reached, or to the fact that the
extent of the numerical errors in the integration of the equations
is higher than that obtained in the case of symmetric transients,
which last only a few dozen time scales. Note that this
satisfactory agreement is observed by using arbitrary initial
conditions in terms of elements of the trigonometrical Schauder
basis for the $L^2$ space, and not by considering as initial
condition the most unstable waves given by the Orr-Sommerfeld
dispersion relation. Moreover, a maximum of the perturbation
energy (in terms of $r$) is found around $k = 0.8$ and confirmed
by both the analyses.

As shown in fig. \ref{IVP_comparison_NMT_IVP_kfixed}, we have also
contrasted our results with the laboratory experimental results
obtained in 1989 by Williamson\cite{W89}, who gave a quantitative
determination of the Strouhal number and wavelength of the vortex
shedding -- oblique and parallel modes -- of a circular cylinder
at low Reynolds number. The comparison is quantitatively good,
because it shows that a wave number close to the wave number that
theoretically has the maximum growth rate at  $Re = 50$ (see part
a of fig. \ref{IVP_comparison_NMT_IVP_kfixed})  has a --
theoretically deduced -- frequency which is very close to the
frequency measured in the laboratory. At this point,  also the
laboratory experimental uncertainty, globally of the order of a
$\pm 10\%$ in an accurate measurement set up,  should be
introduced. The uncertainty associated to the laboratory method
and to the theoretical model (estimated through the difference
between the position of the maximum growth rate showed by the two
cases of asymmetric and symmetric perturbation) overlaps, which
confirms the quality of this comparison. The same quantitative
agreement is observed also at $Re=100$.


\section{Conclusions}\label{Conclusions}

The three-dimensional stability analysis  of the intermediate
asymptotics of the 2D bluff-body viscous growing wake was
considered as an initial-value problem. The velocity-vorticity
formulation was used. The perturbative equations are
Laplace-Fourier transformed in the plane normal to the growing
basic flow. The Laplace transform allows for the use of a damped
perturbation in the streamwise direction as initial condition. In
this regard, the introduction of the imaginary part of the
longitudinal wavenumber (the spatial damping rate) was done to
explicitly include in the perturbation, which otherwise would have
been longitudinally homogeneous, a degree of freedom associated to
the spatial evolution of the system.

An important point is that the vorticity-velocity formulation,
Fourier-Laplace transformation, allows (over a reasonable lapse of computing time) for the following of the
temporal evolution over hundred of basic flow time scales and thus
to observe very long transients.  Such a limiting behaviour  would not  have been
so easily reached by means of the direct numerical integration of the
linearized governing equations of the motion.


Two main transient scenarios have been observed in the region of
the wake where the entrainment is present, the region in between
$x_0=10$ diameters (intermediate)  and $x_0=50$ diameters (far
field), for a $Re$ number equal to $50$ and $100$. A long
transient, where an initial growth smoothly levels off and is
followed either by an ultimate damping or by a slow amplification
for both oblique or 2D waves, and a short transient where the
growth or the damping is monotone. The most important parameters
affecting these configurations are the angle of obliquity, the
symmetry, the polar wavenumber and the spatial damping rate. While
the symmetry of the disturbance is remarkably influencing the
transient behaviour leaving inalterate the asymptotic fate, a
variation of the obliquity, of the polar wavenumber and of the
spatial damping rate can significantly change the early trend as
well as the final stability configuration. Interesting phenomena
are observed. A first one is that, in the case of asymmetric
longitudinal or oblique perturbations,  the system exhibits two
periodic patterns, a first, of transient nature, and a second one
of asymptotic nature. A second phenomenon is that, in the case of
an orthogonal perturbation, albeit always initially set equal to
zero, the transversal vorticity component is the vorticity
component which grows faster. The third  phenomenology  is linked
to the magnitude of the spatial damping rate. Perturbations that
are more rapidly damped in space lead to a larger growth in time.

For disturbances aligned with the flow, the asymptotic behaviour
is shown to be in excellent agreement with the zero order results
of spatio-temporal multiscale modal analyses and with the
laboratory determined frequency and wave length of the parallel
vortex shedding at $Re=50$ and $100$. It should be noted that the
agreement between the initial value problem results and the normal
mode theory is obtained not using as initial condition the most
unstable wave given by the Orr-Sommerfeld dispersion relation at
any section of the wake, but arbitrary initial conditions in terms
of elements of the trigonometrical Schauder basis for the $L^2$
space.

\section*{Acknowledgement}

The authors would like to recognize Robert Breidenthal for his insightful
interpretation of the perturbation dynamics  in the study for wake vorticity.
His experience in this field was a significant contribution.


\end{document}